\documentclass[preprint,11pt]{elsarticle}

\usepackage{lineno,hyperref}
\modulolinenumbers[5]

\journal{New Astronomy Reviews}

\def\Al{$^{26\!}$Al }
\def\Fe{$^{60\!}$Fe }
\def\about{$\sim$}

\def\Msol{$M_\odot$}

\begin{document}
\begin{frontmatter}

\title{Steady-state nucleosynthesis throughout the Galaxy}

\author{Roland Diehl\footnote{Email: rod@mpe.mpg.de}}
\author{Martin G.H. Krause\footnote{current affiliation: Centre for Astrophysics Research, University of Hertfordshire, Hertfordshire, UK}}
\author{Karsten Kretschmer\footnote{current affiliation: DLR Oberpfaffenhofen, Germany}}
\author{Michael Lang\footnote{current affiliation: Airbus Defense and Space, Taufkirchen, Germany}}
\author{Moritz M.M. Pleintinger}
\author{Thomas Siegert\footnote{current affiliation: University of California, San Diego, CA, USA}}
\author{Wei Wang\footnote{current affiliation: School of Physics and Technology, Wuhan University, Wuhan, China}}
\address{Max Planck Institut f\"ur extraterrestrische Physik, D-85748 Garching, Germany}
\author{Laurent Bouchet}
\author{Pierrick Martin}
\address{Institut de Recherche en Astrophysique et Plan\'etologie, 31028 Toulouse, France}

\begin{abstract}
The measurement and astrophysical interpretation of characteristic $\gamma$-ray lines from nucleosynthesis was one of the prominent science goals of the INTEGRAL mission and in particular its spectrometer SPI. Emission from \Al and from \Fe decay lines,  due to their My decay times, originates from accumulated ejecta of nucleosynthesis sources, and appears diffuse in nature.  \Al and \Fe are believed to originate mostly from massive star clusters. The radioactive decay $\gamma$-ray observations open an interesting window to trace the fate and flow of nucleosynthesis ejecta, after they have left the immediate sources and their birth sites, and on their path to mix with ambient interstellar gas. 
The \Al emission image obtained with INTEGRAL confirms earlier findings of clumpiness and an extent along the entire plane of the Galaxy, supporting its origin from massive-star groups. INTEGRAL spectroscopy resolved the line and found Doppler broadenings and systematic shifts with longitude, originating from large-scale galactic rotation. But an excess velocity of $\sim$200~km~s$^{-1}$ suggests that \Al decays preferentially within large superbubbles that extend in forward directions between spiral arms.
The detection of \Al line emission from the nearby Orion clusters in the Eridanus superbubble supports this interpretation. 
Positrons from $\beta^+$decays of \Al and other nucleosynthesis ejecta have been found to not explain the morphology of positron annihilation $\gamma$-rays at 511~keV that have been measured by INTEGRAL. The \Fe signal measured by INTEGRAL is diffuse but too weak for an imaging interpretation, an origin from point-like/concentrated sources is excluded. The \Al/\Fe ratio is constrained to a range 0.2 -- 0.4. Beyond improving precision of these results, diffuse nucleosynthesis contributions from novae (through $^{22}$Na radioactivity) and from past neutron star mergers in our Galaxy (from r-process radioactivity) are exciting new prospects for the remaining mission extensions. 
\end{abstract}


\end{frontmatter}


\section{The $^{26}$Al isotope}

\begin{figure}
\centerline{
  \includegraphics[width=0.8\textwidth]{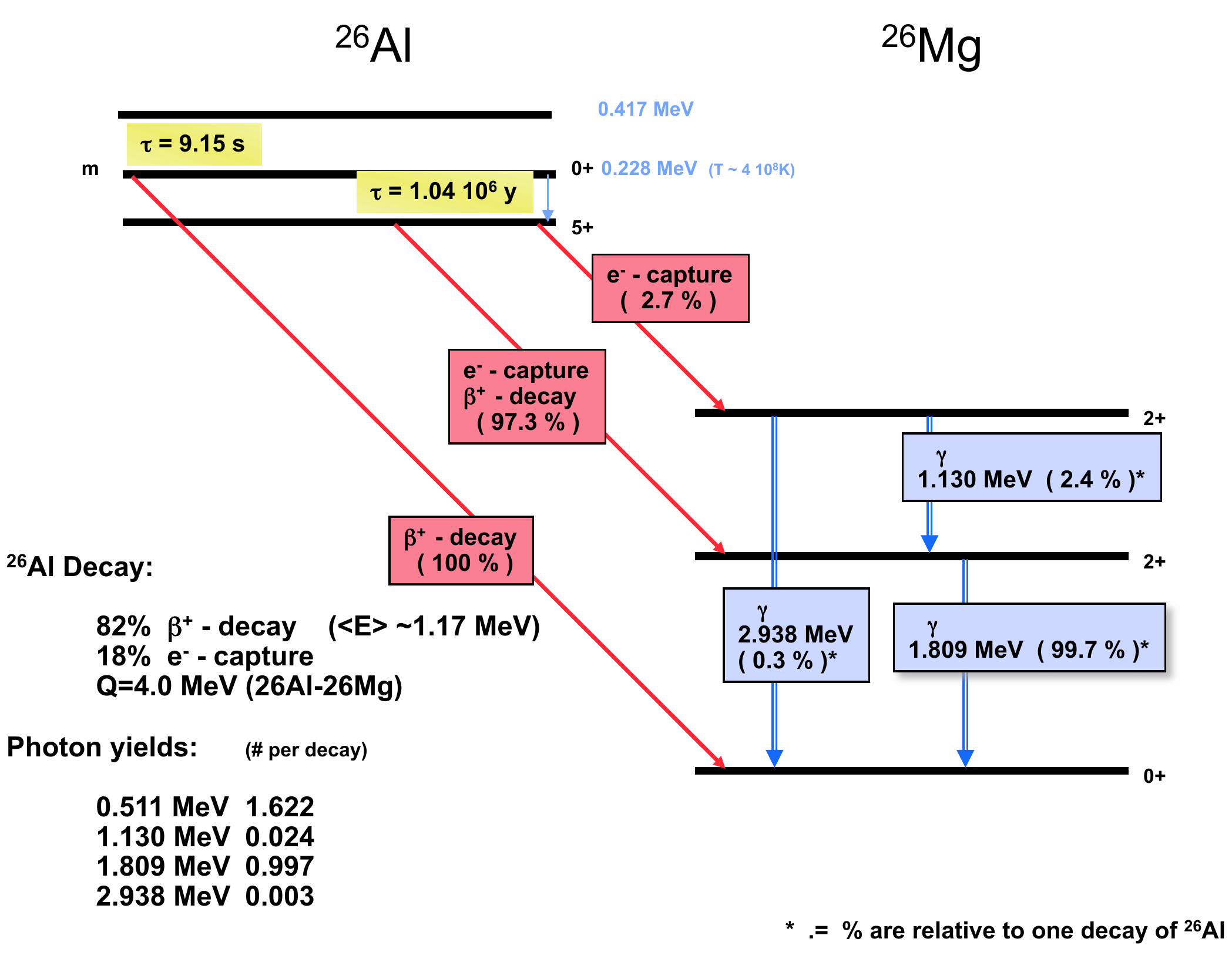}}
  \centerline{
 \includegraphics[width=0.8\textwidth]{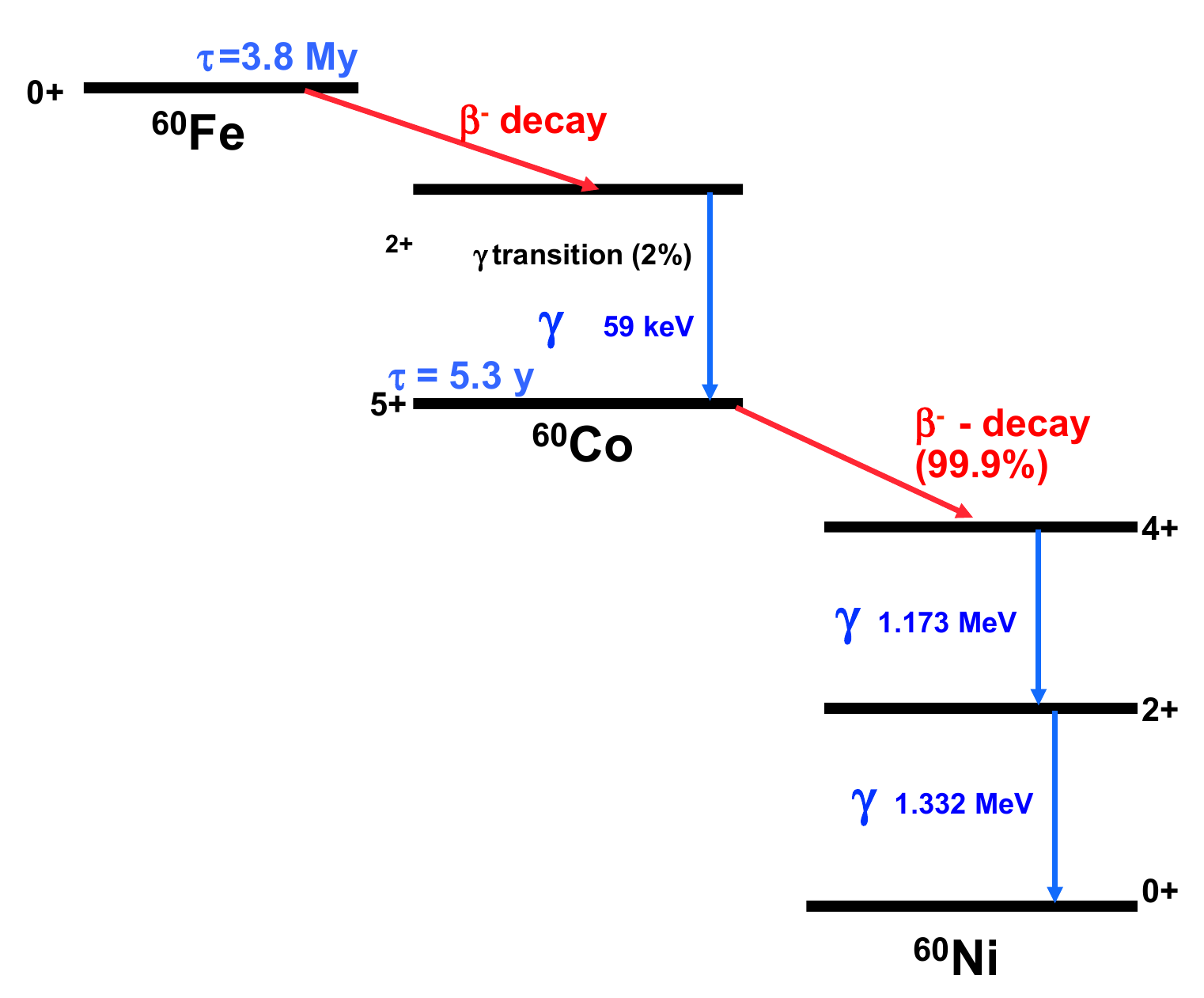}}
  \caption{Radioactive-decay schemes for \Al (top) and \Fe (bottom). The \Al nucleus ground state has a long radioactive lifetime, due to the large spin difference of its state to lower-lying states of the daughter nucleus $^{26}$Mg. \index{isotopes!26Al} An isomeric excited state of \Al exists at 228 keV excitation energy. If thermally excited, \Al may decay through this state. Secondary products, lifetime, and radioactive energy available for deposits and observation depend on the environment. 
  The \Fe isotope decays with a radioactive lifetime of 3.8~My through $^{60}$Co to $^{60}$Ni. Note that per decay, two $\gamma$-ray photons are obtained.}
  \label{fig_26Al60Fe-decays}
\end{figure}

Radioactive decay of the \Al nucleus (see Fig.~\ref{fig_26Al60Fe-decays}) provides a special opportunity to astrophysics, as this unstable isotope is unusually long-lived ($\tau$=1.04~My) among the abundantly-produced lighter nuclei up to the iron group. 
This long radioactive lifetime of \Al is due to special nuclear levels arising from its odd-odd composition of 13 protons and 13 neutrons, leading to a high threshold for $\beta$ transitions from this particular shell configuration of neutrons and protons. 
The ground state of \Al is an (unusually-high) $5+$ spin state, while lower-lying states of the neighbouring isotope $^{26}$Mg have states $2+$ and $0+$, so that a rather large change of angular momentum  must be carried by radioactive-decay secondaries. This causes the large $\beta$-decay lifetime of \Al.
Excited states of \Al at energies 228, 417, and 1058~keV with $0+$ and $3+$ spins are more favourable for decay due to their smaller angular momentum differences to the $^{26}$Mg states. This leads to a lifetime of 9.15~s for the $0+$ \emph{metastable} state $^{26m}$Al, which could be thermally excited in nucleosynthesis sites exceeding a few 10$^8$K. 
\Al thus may decay via this state without any $\gamma$-ray emission. 

\Al is an excellent and promising tracer for stellar nucleosynthesis, as it survives long enough after production inside a star or explosion to have a good chance to decay only after being ejected into the interstellar medium. Additionally, it serves as a thermometer for its production environment, hence also provides astrophysical diagnostics for its production site. Moreover, due to the long radioactive lifetime, decay and $\gamma$-ray emission occur millions of years after nucleosynthesis, thus tracing the fate and flow of ejecta after supernova explosions.
Note that individual supernova remnants can only be observed  in radio or X rays up to ages of several 10~ky after a supernova explosion.

Candidate cosmic sources of \Al are all environments of H burning, as pre-existing $^{25}$Mg converts to \Al through the $^{25}$Mg(p,$\gamma$)$^{26}$Al reaction.
Nova explosions first come to mind, being explosive hydrogen  burning sites on the surface of a white dwarf resulting from stellar evolution and hence likely including $^{25}$Mg (or else, producing it from more-abundant $^{24}$Mg through hydrogen burning, $^{24}$Mg(p,$\gamma$)$^{25}$Mg ). Novae thus were considered the prime candidate for \Al production \citep{Clayton:1993}.
In principle, hydrogen burning occurs in all stars during the main sequence.
Stars of high masses above 10--25~\Msol\  evolve rather rapidly, within $\sim$10~My through the entire main sequence \citep{Schaller:1992}.
Thus, a significant part of the \Al produced in this phase will still be alive while the star enters the giant phase with core helium burning and shell hydrogen burning. This is accompanied with substantial structural changes of the star, leading to significant outward mixing of hydrogen-burning ashes into the envelope, and also to mass loss with strong stellar winds.
Thus, \Al may be carried away with the wind, in particular when a star enters the Wolf-Rayet phase \citep{Signore:1993,Meynet:1997}, which is thought to occur for stars with masses above $\sim$25~\Msol\  at solar metallicity. 
Such massive stars then end their evolution soon after the giant transition as core-collapses, likely producing supernova explosions, at least for masses below $\sim$40~\Msol.
The supernova now will eject the envelope, including all \Al that may still reside therein, and enhanced by some additional \Al nucleosynthesis from explosive burning as the supernova shock rushed through the star. 
Also massive stars that are below the minimum mass to evolve towards a core-collapse supernova, i.e. AGB stars of the mass range 3--8~\Msol\  have been discussed as candidate \Al sources \citep{Bazan:1993,Forestini:1997,Karakas:2007}. Here, in particular, nuclear burning at the hot inner edge of the hydrogen-burning shell provides a suitable environment for \Al synthesis. 
Massive stars in general, as well as their supernovae, are thus plausible sources of \Al, from an extended phase during their evolution, and eventually from the terminal core-collapse supernova.
Predictions were derived from theoretical modeling for the entire \Al amount present in the Galaxy from each of these source types. 
Supernovae from core collapse of massive stars were thought to contribute 2.1$\pm$1.1~\Msol\  \citep{Timmes:1995}. 
The earlier Wolf-Rayet phase was estimated to contribute a galactic total of 0.5 \Msol\  with a factor $\sim$3 uncertainty \citep[see][and references therein]{Prantzos:1996a}.   
Models for these two candidate sources were best established, while models for novae and AGB stars are more uncertain, by comparison. 
Estimates for \Al from novae range from 0.1 to 5~\Msol, with large uncertainties, mainly from a lack of progenitor knowledge, and from lacking still an also quantitatively-realistic nova model. From more-recent models, maybe 0.2~\Msol\ of Galactic \Al may be due to classical novae, with again a factor $\sim$3 uncertainty \citep[see][for a description of nova nucleosynthesis]{Jose:1998}. 
The AGB star contribution is now estimated \citep{Karakas:2014} to be below 10\% of that of massive stars. 
Note that both AGB stars and novae are clearly identified as \Al producers from interstellar-grain inclusions in meteorites \citep[see][for a review]{Clayton:2004}. Both these two types of sources are copious producers of dust grains, more so than supernovae or WR stars; interstellar-grain samples are biased towards measuring \Al from such dust-producing sources.

\subsection{Pre-INTEGRAL measurements of \Al}
The first report of a $\gamma$-ray line from a radioactive nucleus of cosmic origin arose from the HEAO-C mission 1979/1980 measurements. This satellite carried a Ge detector with high spectral resolution, so that a line at 1809~keV could plausibly be attributed it to decay of live \Al in the Galaxy's interstellar medium \citep{Mahoney:1984}. 
\Al radioactive decay within 1~My requires a source within the past few million years, which is a rather brief recent time span in the Galaxy's history, which spans 12~Gy or more. Therefore this is proof of currently-ongoing nucleosynthesis, observed most-directly through the characteristic $\gamma$-ray emission that is a consequence of the decay of this radioactive nucleus\footnote{Cosmic-ray activation of this nuclear line in $^{26}$Mg would also produce other, brighter, lines from more abundant species, which are not observed.}.

\begin{figure}  
\centering 
\includegraphics[width=1.0\textwidth]{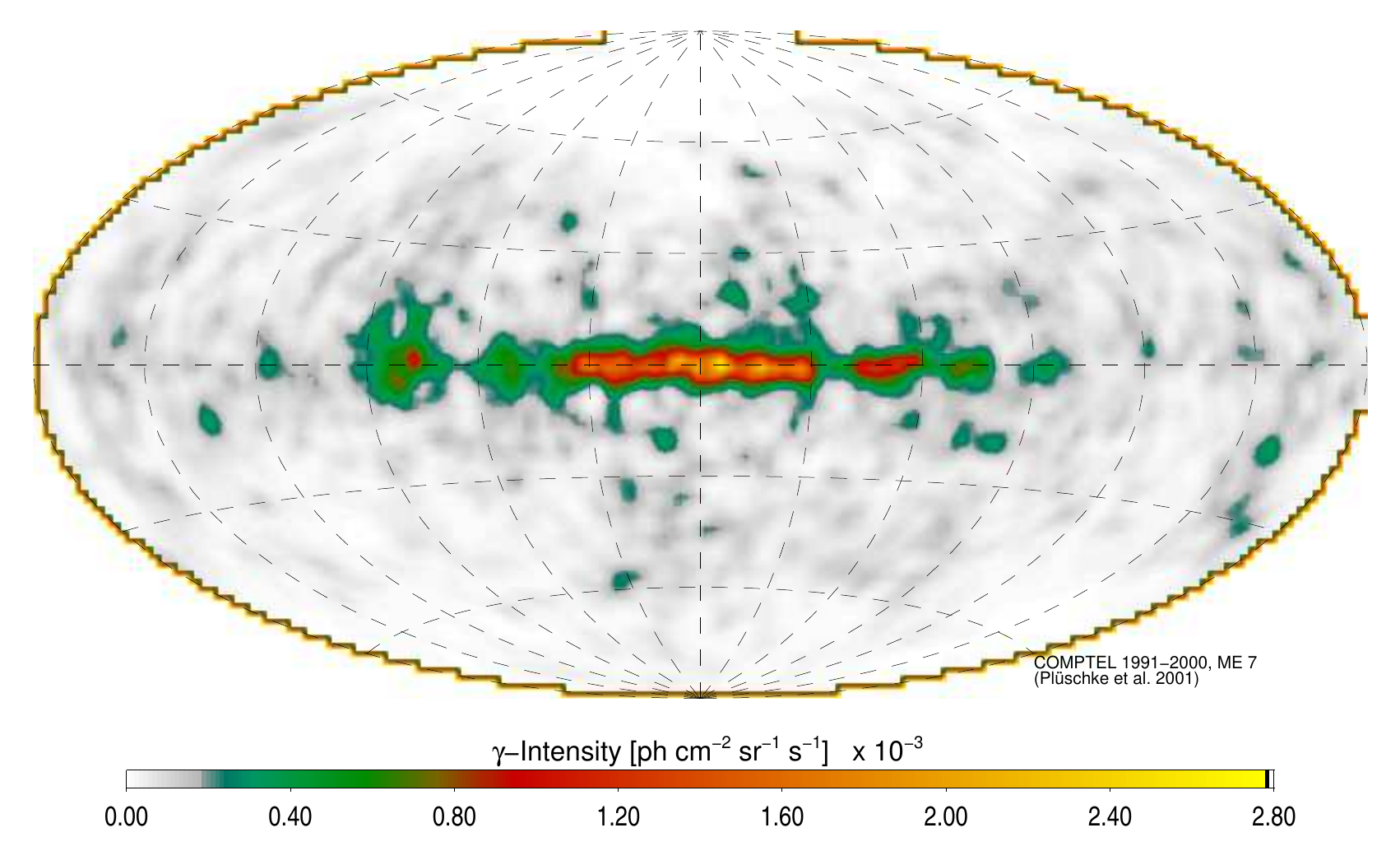} 
\includegraphics[width=1.0\textwidth]{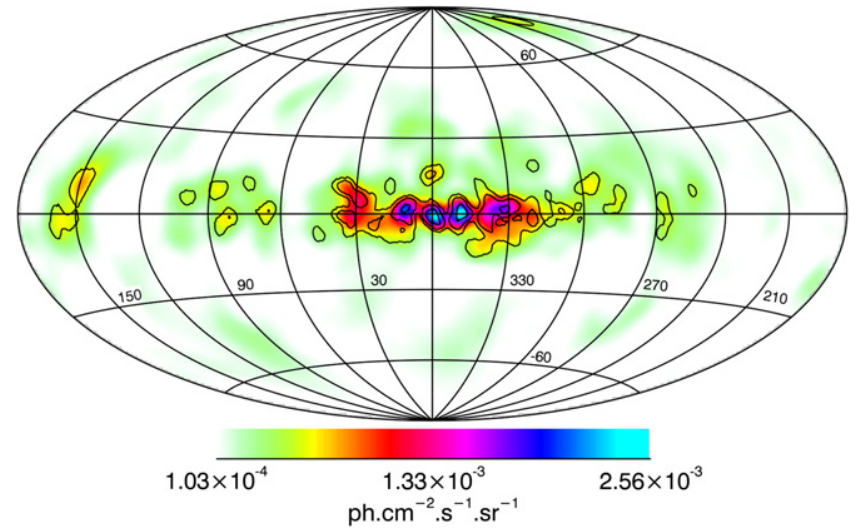}
\caption{\emph{Top:} The \Al sky as deconvolved from the COMPTEL data,  using a maximum-entropy method  \citep{Pluschke:2001c}. 
\emph{Bottom:} The \Al sky as deconvolved from INTEGRAL's sky survey from ten years of data with the SPI imaging spectrometer, also using the maximum-entropy imaging method \citep{Bouchet:2015}.}
\label{fig:almaps} 
\end{figure}   

The COMPTEL sky survey, accumulated over nine years, then provided a sky image in the \Al $\gamma$-ray line  \citep{Diehl:1995b,Oberlack:1996,Pluschke:2001c} (Fig.~\ref{fig:almaps}, top). This effectively is a map of nucleosynthesis activity in our current Galaxy, with an effective exposure of the recent few million years due to \Al decay. 
The \Al maps that were obtained from COMPTEL event measurements through different varieties of iterative deconvolution methods all showed structured \Al emission, though in detail at different degrees depending on methods and sky exposure. 
All images presented extended along the plane of the Galaxy \citep{Oberlack:1996,Pluschke:2001c,Knodlseder:1999,Knodlseder:1999c,Strong:1999a}, in broad agreement with earlier expectations of \Al being produced throughout the Galaxy and mostly from massive stars and their supernovae \citep{Lingenfelter:1978,Prantzos:1993}.

The irregular distribution of \Al emission all along the plane of the Galaxy provided a main argument for the consensus that massive stars dominate the production of \Al \citep{Prantzos:1996a}. 
Massive stars have been recognised to preferentially occur in clusters \citep{Lada:2003,Lada:2005}. The star formation regions also would be expected to be located near spiral arms, where hierarchical gravitational collapse and turbulence are efficient in creating new stars, and feedback from massive stars shapes surrounding gas \citep{Dobbs:2014a,Reid:2014}.
This plausibly explains structured emission along the plane of the Galaxy. 
Prominent more-nearby massive-star regions appear particularly bright in \Al emission, such as the Cygnus region (see below), which appears in all image variants.
Compared to this, \Al ejections from novae or intermediate-mass AGB stars would appear with a much smoother spatial distribution \citep{Clayton:1993,Bazan:1993}, due to the much larger number of individually-contributing sources with their individually-lower \Al yields.  

$^{26}$Al emission from prominent and particularly-bright regions had been discussed in theoretical work. 
The science goal for such individual source regions was to make use of the knowledge about the stellar content, in particular the age estimates that can be obtained from the census of the remaining population of (high mass) stars that can still be observed, i.e. that had not  already ended their evolution in a gravitational collapse. 
Population synthesis studies and their astrophysical potential in the context of \Al measurements had been first discussed in detail by Cervino et al. (2001), and later by Voss et al. (2009). 
The first convincing association of a \Al $\gamma$-ray signal with a specific massive-star region was presented for the Cygnus region, based on COMPTEL data \citep{del-Rio:1996,Knodlseder:2000, Pluschke:2000}. The Cygnus OB2 association in particular was recognised as one of the most-massive star groups in our Galaxy (Knoedlseder et al. 1999).
Interpretations of \Al emission in the region along the entire line of sight towards Cygnus include about 9 stellar associations along the line of sight and from cavities of different sizes \citep{Comeron:1994a,Pluschke:2001a,Pluschke:2002}.  
Hints of \Al emission for other nearby massive-star regions also had been derived from COMPTEL measurements: Most prominently, emissions were discussed from the Orion region at \about~400~pc distance, and the Vela region with candidate sources at distances from \about~200 out to 1500~pc \citep{Diehl:2002}.

A Galaxy-wide interpretation of the \Al $\gamma$-ray measurements is attractive, as the total mass of \Al seen in observations can be confronted with theoretical estimates of the amounts due to the entire Galactic population of sources of a particular type of candidate source. 
The population of sources can be estimated from their link to the Galaxy's star formation rate, and a spatial distribution that reflects the evolution in its simplest form. 
Such analysis needs to resolve the distance uncertainty, when assigning a measured flux along a line of sight to the brightness of particular source regions. Localised regions along the line of sight which may efficiently produce \Al  need to be accounted for properly, using other astronomical knowledge.
The massive star census in the Galaxy is well known from astronomical measurements of their thermal as well as maser emission, the former only out to distances of a few kpc. Beyond, many regions of the Galaxy are occulted for direct measurements. Therefore, one is left with some uncertainty about their Galaxy-wide distribution. Probably, the \emph{molecular ring} around the center of our Galaxy at a radial distance of 3--4~kpc from the center is a prominent birth site for massive stars, as are molecular clouds swept up along the Galaxy's spiral arms \citep{Dame:2001,Reid:2014}. 
The  total amount of \Al in the Galaxy can be estimated from the measured $\gamma$-ray flux, using as a plausible assumption for the distances to the emission regions a galaxy-wide distribution. 
Based on COMPTEL data, an \Al amount around 3~\Msol\ had been advertised \citep{Diehl:1995b,Knodlseder:1996}. Under plausible standard assumptions about the stellar population and their evolution, this converts into a Galactic star formation rate of $\sim$5~\Msol~y$^{-1}$, or a core-collapse supernova rate of 3--4 per century \citep{Timmes:1997a}. \citep[For a detailed discussion of such conversions, see][]{Diehl:2006d}.

The GRIS balloon experiment, carrying high-resolution Ge detectors, had provided an indication that the \Al line was significantly broadened beyond instrumental resolution to 6.4~keV (FWHM) \citep{Naya:1996}. This implied kinematic Doppler broadening of astrophysical origin, with velocities of 540~km~s$^{-1}$. Considering the $1.04$~My decay time of $^{26}$Al, such a large line width would naively translate into kpc-sized cavities around \Al sources, or major fractions of  \Al to be condensed on grains that could retain their momentum in a ballistic flow through tenuous interstellar gas \citep{Chen:1997,Sturner:1999}. 
COMPTEL's scintillation $\gamma$-ray detectors lacked the spectral resolution required for line identification and spectroscopic studies: It had \about~200~keV instrumental resolution (FWHM), compared to \about~3~keV for Ge detectors, at the energy of the \Al line. 
Therefore, another measurement of the \Al line at high spectral resolution was also important to place these tantalizing hints on firmer footing.

\subsection{INTEGRAL measurements and lessons} 
The INTEGRAL space observatory \index{INTEGRAL} with its Ge-detector based spectrometer SPI, launched in 2002, provided a wealth of high-quality spectroscopic data accumulated over its long mission lifetime. 
This allowed a first measurement of kinematic data about \Al \citep{Kretschmer:2013} and a higher-precision study of the Galaxy-wide \Al aspects \citep{Diehl:2006c,Diehl:2006d} and their role to understand stellar feedback \citep{Krause:2015}.

An all-sky image had been constructed from 10 years of data \citep{Bouchet:2015}, and is shown in Fig.~\ref{fig:almaps} (lower graph) together with the COMPTEL image. 
The sky exposure during the INTEGRAL mission was largely concentrated towards the inner Galaxy and their X-ray binaries, so that any signals and features in the outer Galaxy would be seen with less significance.
Any such image is potentially susceptible to systematics and uncertainties from the instrument, due to statistical noise (Poisson) and a response that cannot be directly inverted, so that forward convolution methods are applied. But due to different imaging methods and detectors, as well as backgrounds, systematics effects will be different among instruments, making comparisons a valuable assessment of the astronomical imaging results. 
The total background count rate of a coded-mask instrument is much higher than for the Compton telescopes. On the other hand, the imaging principle is more straightforward, with a single interaction and detector type. The \Al images from INTEGRAL and COMPTEL show striking similarities, and several features of the COMPTEL-based \Al image re-appear in the INTEGRAL result (see Fig.~\ref{fig:almaps}).
In view of the different ways those images were obtained, we consider as confirmed that there is diffuse emission all along the plane of the Galaxy, with emission peaks or hot spots reminiscent of known massive star groups. 

\subsubsection{The mass of \Al in the Galaxy}

\begin{figure}  
\centerline{
\includegraphics[width=0.67\textwidth]{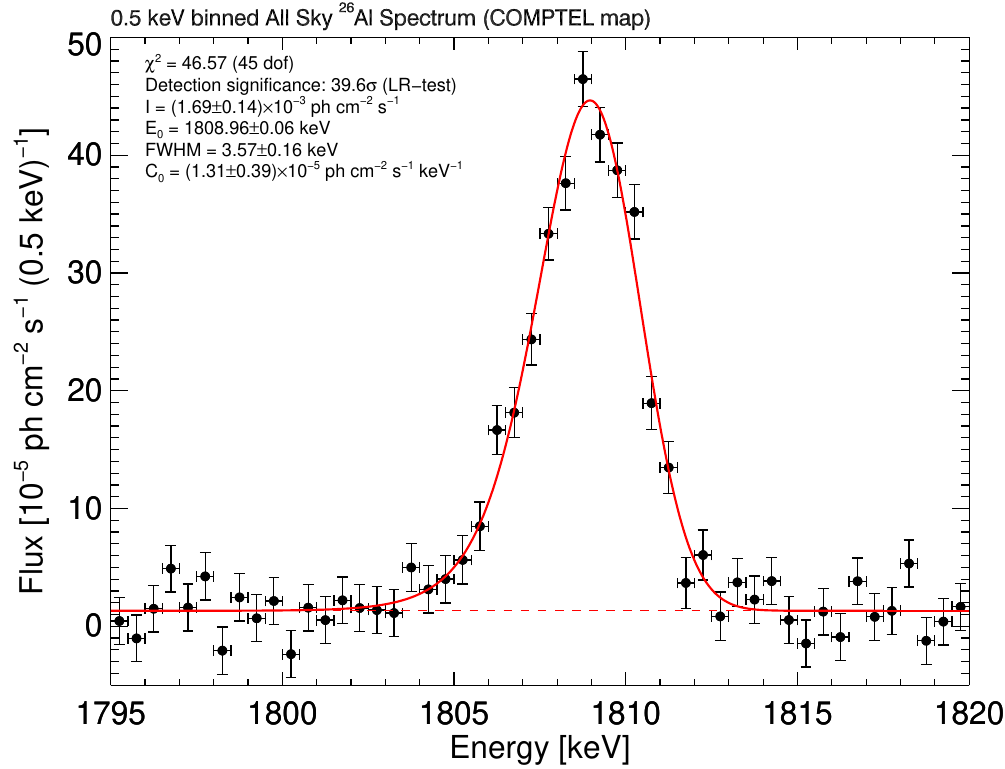}
\includegraphics[width=0.33\textwidth]{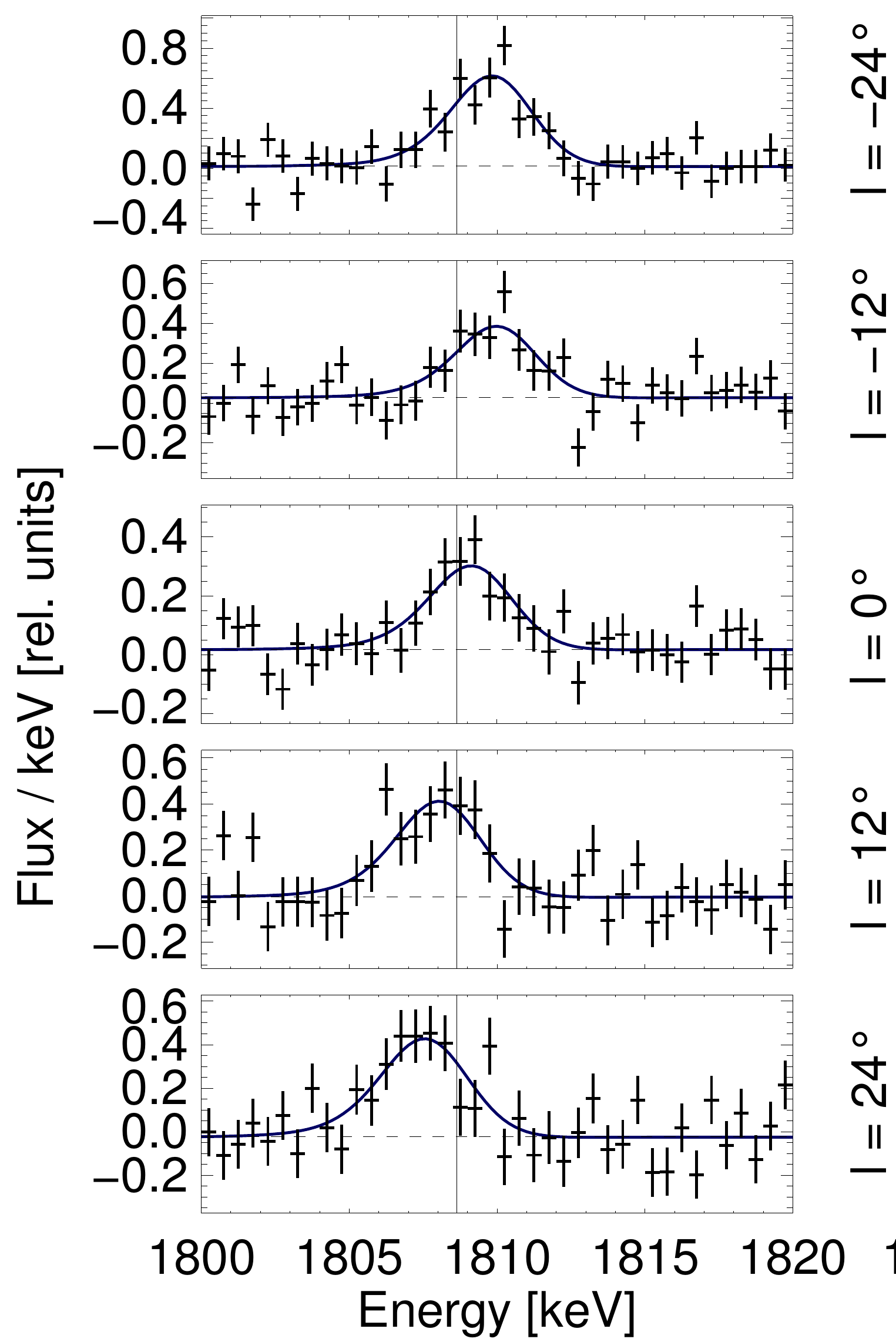}}
\caption{The \Al line as seen with INTEGRAL's high-resolution spectrometer SPI and 13 years of measurements integrated \citep{Siegert:2017} {\it(left)}. 
\Al line variation with Galactic longitude\citep{Diehl:2006d} {\it (right)}.  This shift of the line centroid reflects the kinematics of \Al towards the inner Galaxy in INTEGRAL/SPI measurements 
 \citep{Kretschmer:2013}.}
\label{fig:almap-spec} 
\end{figure}   

\begin{figure}  
\centerline{ 
\includegraphics[width=1.0\textwidth]{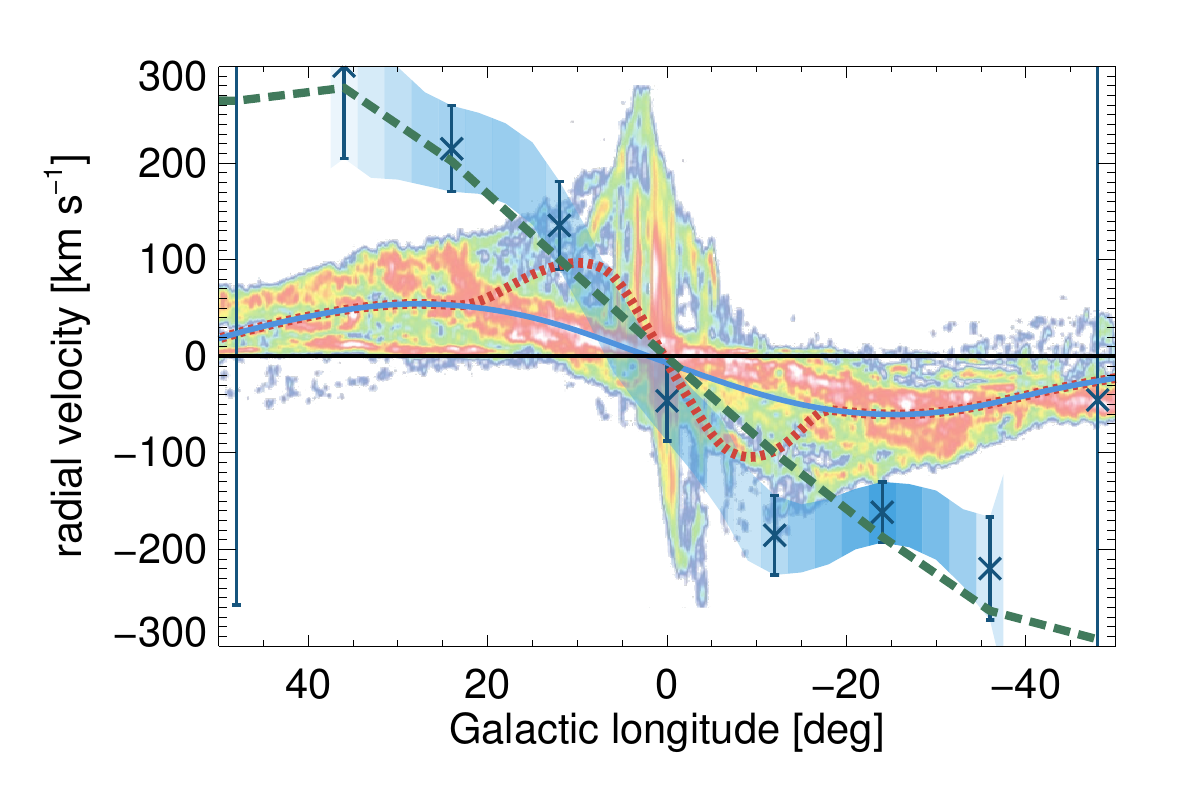}}\centerline{ \includegraphics[width=0.6\textwidth]{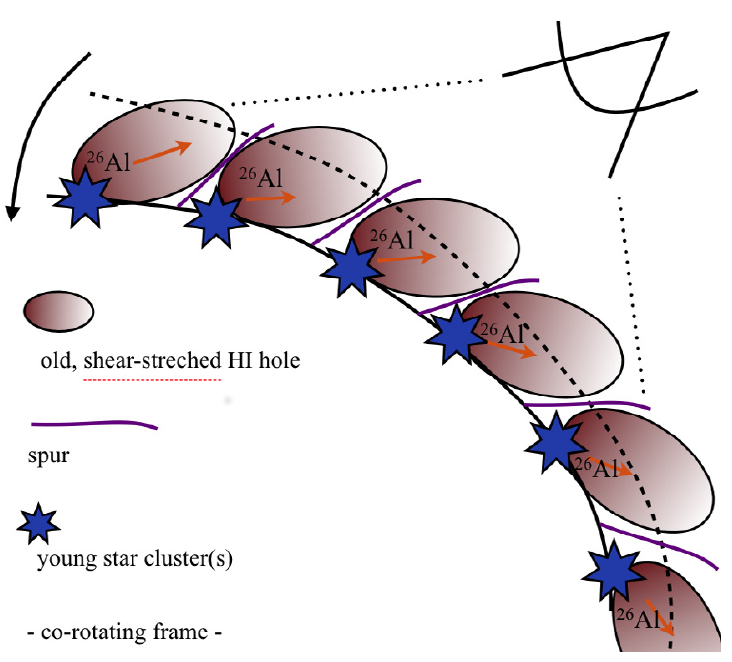}}
\caption{{\it (Top:)} Kinematics of \Al towards the inner Galaxy, from INTEGRAL/SPI measurements. This longitude-velocity diagram for hot ISM as traced through \Al in the inner Galaxy shows the trend from the Galaxy's large-scale rotation. The underlying color plots show the corresponding kinematics from molecular gas as traced through CO data. The \Al traced hot gas shows systematically higher velocities by about 200 km~s$^{-1}$ in the direction of Galactic rotation. 
 {\it(Bottom)}:  Scenario for asymmetric surroundings of \Al sources. At time of nucleosynthesis product ejection, massive star groups could be located at the leading edges of spiral arms, thus presenting more material moving away from spiral arms at higher velocities than moving 'backwards' into higher density regions \citep{Krause:2015}.
 \citep{Kretschmer:2013}}
\label{fig:al_rotation-model} 
\end{figure}   

The integrated flux of \Al $\gamma$-rays provides a measurement of the total mass of \Al in our Galaxy.
 INTEGRAL/SPI data were used to obtain a value of (2.8 $\pm$0.8)~\Msol\  of \Al  \citep{Diehl:2006d}, in a study comparing many alternative views of massive star activity in the Galaxy.
A systematic limitation of any galaxy-wide parameter determination occurs from potential bias of the specific location of the Sun within the Galaxy and with respect to spiral arms. Gamma rays are at an advantage here due to their penetrating nature, reaching us also from sources of the distant regions of our Galaxy. However, there may be sources in the vicinity of our vantage point in the solar system, 8~kpc from the Galaxy's center. 
An account for the more-nearby sources (discussed in the next subsection) leads to a reduction of the Galaxy-wide \Al content. Successively accounting for foreground sources as they could be discriminated, improved values were determined. Therefore, the \Al mass determined as such from the INTEGRAL data directly range from 1.7$\pm$0.2~\Msol \citep{Martin:2009} to 2.0$\pm$0.3~\Msol \citep{Diehl:2010,Diehl:2016b}; the quoted uncertainty includes estimates of residual systematics from such foreground.
Alternatively, a bottom-up modelling of the INTEGRAL-observed \Al sky from massive-star clusters \citep{Pleintinger:2020} has shown that this quoted uncertainty estimate appears realistic. 
A large-scale simulation had indicated that the \Al mass in the Galaxy could be even lower \citep{Fujimoto:2018}.
On the other hand, other Galactic-scale hydrodynamic simulations with \Al sources distributed along spiral arms reproduce the INTEGRAL observations with a steady-state mass of $\sim$4~\Msol, values between 2.5 and 6~\Msol\ being consistent with these simulations \citep{Rodgers-Lee:2019}. It remains to be seen how morphology improvements from imaging as well as simulation work will refine the role of nearby and intermediate-distance clusters.

With a mass estimate from $gamma$-ray data, we can use the theoretical yield estimates per mass and integrate over the mass distribution function to obtain the normalization factor of the Galactic star formation or core-collapse supernova rate. This is an important independent way to derive this key quantity for our Galaxy, because it derives from a measured star formation tracer that is unaffected by occultation and thus galaxy-wide, in contrast to, e.g.,OB or WR star counts\citep[see][for a discussion of measurement alternatives]{Diehl:2006d,Chomiuk:2011}. From INTEGRAL/SPI \Al measurements, a value of 1.9$\pm$1.1~SN~century$^{-1}$ has been estimated \citep{Diehl:2006d}.  Correspondingly scaled down due to the above-discussed foreground sources, this obtains 1.3$\pm$0.4~SN~century$^{-1}$ \citep{Diehl:2016b} .

\subsubsection{\Al decay within superbubbles}
One of the main achievements in INTEGRAL/SPI \Al measurements is the high-resolution spectroscopy of the \Al line, which resolves the line and allows to analyze its shape (see Fig.~\ref{fig:almap-spec}).
An astrophysical broadening of the all-sky integrated  \Al line was discovered, with a value of 1.4~keV ($\pm$0.3~keV). This is much smaller than what had been reported from the balloon instrument data with a Ge detector \citep{Naya:1996}, and corresponds to 175~km~s$^{-1}$ ($\pm$45~km~s$^{-1}$) in velocity space. 
The line width of the \Al line as seen from the Galaxy in Fig.~\ref{fig:almap-spec} (left) reflects Doppler broadening from the large-scale rotation within the Galaxy, and from peculiar \Al motion.
As shown in Fig.~\ref{fig:almap-spec} (right), there is a clear and systematic trend with longitude, that shows the kinematic signature from large-scale Galactic rotation. This could be clearly detected  \citep{Kretschmer:2013} with sufficient INTEGRAL exposure  \citep[confirming earlier indications  \citep{Diehl:2006d}; see discussion in][for how sufficient detail had been accumulated over the years of the INTEGRAL mission to enable such a measurement]{Diehl:2013b}. 
As shown in the upper graph of Fig.~\ref{fig:al_rotation-model}, 
the bulk velocities seen in the Doppler shifts of the \Al line as viewed towards different Galactic longitudes turn out to be on the order of several 100~km~s$^{-1}$.
This is a much higher velocity than commonly found in cold and dense gas, such as the bulk velocities of molecular clouds traced by carbon monoxide \citep{Dame:2001},
which characterises the large-scale rotation of objects in the Galaxy in general.
Evidently, the velocities measured from \Al kinematics exceed typical Galactic rotation significantly, by up to $\approx$200~km~s$^{-1}$. 
Clearly, the bulk of massive-star ejecta as traced by \Al is not mixed with cold interstellar gas on a time scale $\approx$10$^6$y after their ejection, as otherwise smaller line-of-sight velocities would be expected.

When we consider that \Al will stream into the medium surrounding its sources while decaying, it is clear that the velocity measurement from \Al will be affected by the conditions in the vicinity of the source.  
Ejection from the sources plausibly should be isotropic. Typical ejection velocities  are on the order of 1000~km~s$^{-1}$, both for core-collapse supernovae \citep{Janka:2007a} and for Wolf-Rayet stellar winds \citep{Crowther:2007}. 
If ejecta could travel freely for about a kpc before \Al decay, such high initial velocities would still be reflected in the \Al $\gamma$-ray line.
Structures in the interstellar medium with their characteristic sizes at or below this scale can therefore constrain ejecta expansion in different directions, thus making expansion deviate from isotropic. 

Based on such ideas, an asymmetry scenario was proposed (see sketch in lower graph of Fig.~\ref{fig:al_rotation-model}):  Massive stars inside the Galaxy's co-rotation radius would be formed within spiral arms, but travel towards the leading edges of spiral arms while evolving into their Wolf Rayet phases, and in any case before they explode as supernovae. 
Then, the ejection of nucleosynthesis ejecta would occur in a region that is characterised by a density gradient, higher density surroundings from the spiral arm being located preferentially behind the massive stars, as seen from an observer in the direction of Galactic rotation.
Massive-star ejecta are quickly thermalised at shocks, and fill the superbubbles created from earlier massive-star winds and radiation. 
It is well known that such superbubbles are asymmetric, being shaped by shear, local density gradients, and embedded sources \citep{Baumgartner:2013,Krause:2014,Krause:2018}.
So, ejecta would be decelerated if streaming towards the spiral arms, while streaming at higher velocities would be allowed into the lower-density inter-arm regions. 
Such a scenario  is illustrated in Fig.~\ref{fig:al_rotation-model} \citep{Krause:2015}, placing the candidate \Al sources along inner spiral arms and preferentially onto their inner ends approaching the Galaxy's bar. This results in a longitude-velocity trend as shown by the green-dashed line in Fig.~\ref{fig:al_rotation-model}. 
In these calculations, the pitch angle of a logarithmic spiral structure model has been fitted (and obtains values in agreement with common results inferred otherwise), as well as an ejecta velocity of 200~km~s$^{-1}$, which corresponds to the typical expected sound velocity in superbubbles. 
The significant enhancement of apparent velocities as observed from \Al  therefore  suggests that sources of \Al may be typically surrounded by large interstellar cavities (i.e., superbubbles). 
These large cavities could have been left over from previous stellar generations and their massive-star activity. Or, alternatively, the winds of the most-massive stars of a massive-star group, which evolve on shortest time scales,  enter their Wolf-Rayet phase with strong winds only about 3~My after their formation and thus create such a cavity.
Superbubbles of order $\sim$kpc as typical surroundings of massive stars with ages of few to tens of My then present a new scenario for how new nuclei may be recycled into the general flow of cosmic gas as part of cosmic chemical evolution  \citep{Krause:2015}. 
The sizes of such cavities plausibly extend up to one kpc. 
This exceeds the scale height of the molecular and atomic interstellar medium. Massive star ejecta as traced by \Al therefore may recycle to the next generation of stars on a long way over the halo of the Milky Way on timescales likely exceeding 10~Myrs. 
Recent galaxy-scale hydrodynamics simulations with \Al ejection show that the \Al scale height is sensitive to the assumed galaxy structure, clustering of star formation as well the gas density in the lower halo \citep{Fujimoto:2018,Pleintinger:2019,Rodgers-Lee:2019}. With realistic assumptions, scale heights of about 2~kpc near the solar radius are found \citep{Rodgers-Lee:2019}, similar to values derived from the INTEGRAL data \citep{Pleintinger:2019}.
 The simulations thus support the interpretation of the INTEGRAL data as tracing massive-star ejecta through superbubbles into the Galactic halo.

Ejecta may thus be returned into the ambient interstellar medium 
on time scales beyond 10$^7$~years.  
INTEGRAL observations of \Al have shed new light onto the evolution of enriched gas on such time scales, which are difficult to constrain through observations otherwise.

\begin{figure}  
\centerline{  
\includegraphics[width=0.8\textwidth]{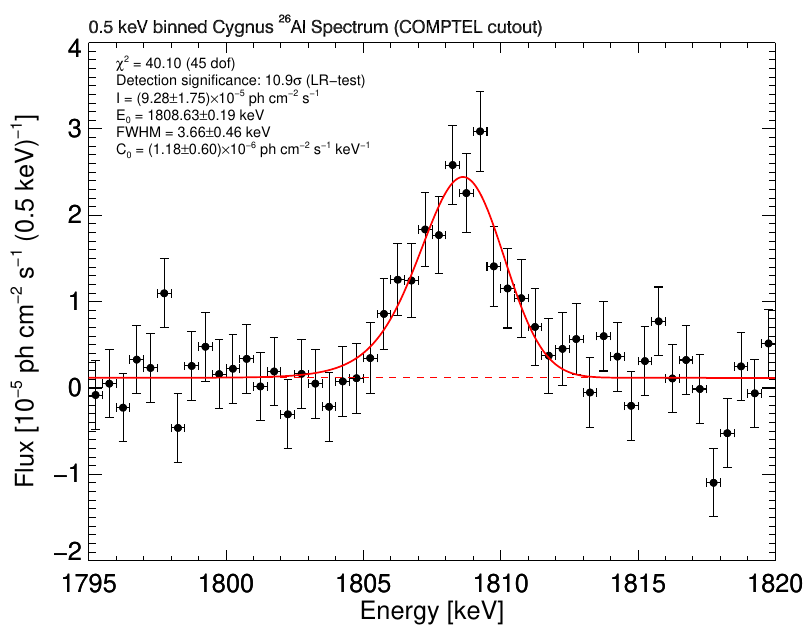}}\centerline{
\includegraphics[width=0.8\textwidth]{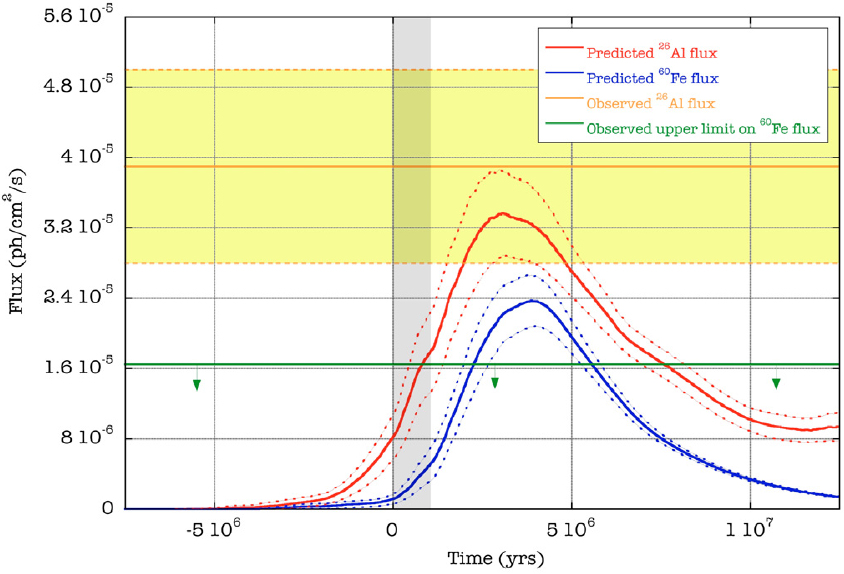}}
\caption{{\it (Top:)} The \Al signal measured from the Cygnus region  \citep{Siegert:2017a}. {\it (Bottom:)} The time history of \Al production in the Cygnus complex, as predicted from population synthesis, is compared to INTEGRAL's early $\gamma$-ray flux measurements\citep{Martin:2010b}. Expectations from such populations synthesis are on the low side of observed \Al $\gamma$-rays, in particular as the lower metallicity compared to solar for the Cygnus region is adopted. The horizontal shaded area presents the range given by the \Al $\gamma$-ray data, the dashed lines bracket the uncertainty range of predictions from recent massive-star models through population synthesis, and the vertical shaded area indicates the current time (from adopted cluster ages; the impact of stellar rotation on age estimates determines the width of the shaded area). \citep[Figure adapted from][]{Martin:2010b}.
}
\label{fig:al_cygnus} 
\end{figure}   

\subsubsection{Specific regions and massive-star groups} 
A detailed test of our understanding of massive-star activity can be made in specific, nearby, and well-constrained massive-star groups, i.e., for which the formation of stars likely is coeval (i.e., due to one single star formation epoch), and the stellar census as well as other astronomical constraints are available. 

The Gould Belt has been recognised as a region of local and nearby stellar associations arranged in an elliptical-belt-like structure\citep{Olano:1982,Poppel:1997,Poppel:2010,Maiz-Apellaniz:2004,Perrot:2003}. Several associations believed to be Gould Belt members also appear to be aligned with \Al emission peaks. 
Interesting results on \Al from these regions were obtained with INTEGRAL for the Cygnus \citep{Knodlseder:2004a,Martin:2009}, Orion \citep{Siegert:2017a}, Vela \citep{Maurin:2004}, Carina \citep{Voss:2012}, and Scorpius-Centaurus \citep{Diehl:2010} groups of massive stars. 

The Cygnus region \citep{Reipurth:2008} has been prominent as an \Al source even in first images of \Al gamma rays \citep{Diehl:1995}. 
A detailed investigation of the Cygnus region and its \Al sources was provided from a PhD thesis \citep{Pluschke:2001a} based on COMPTEL observations, following earlier work \citep{del-Rio:1996}. 
INTEGRAL observations confirmed the prominent \Al emission from the Cygnus region \citep{Knodlseder:2004a,Martin:2010b}, with indications of a slightly-broadened \Al line pointing to \Al decay from a hot medium (Doppler broadening). 
Fig.~\ref{fig:al_cygnus} (left) shows the \Al spectrum after 13 years of cumulative exposure \citep{Siegert:2017}. 

\begin{figure}  
\centerline{ 
\includegraphics[width=0.9\textwidth]{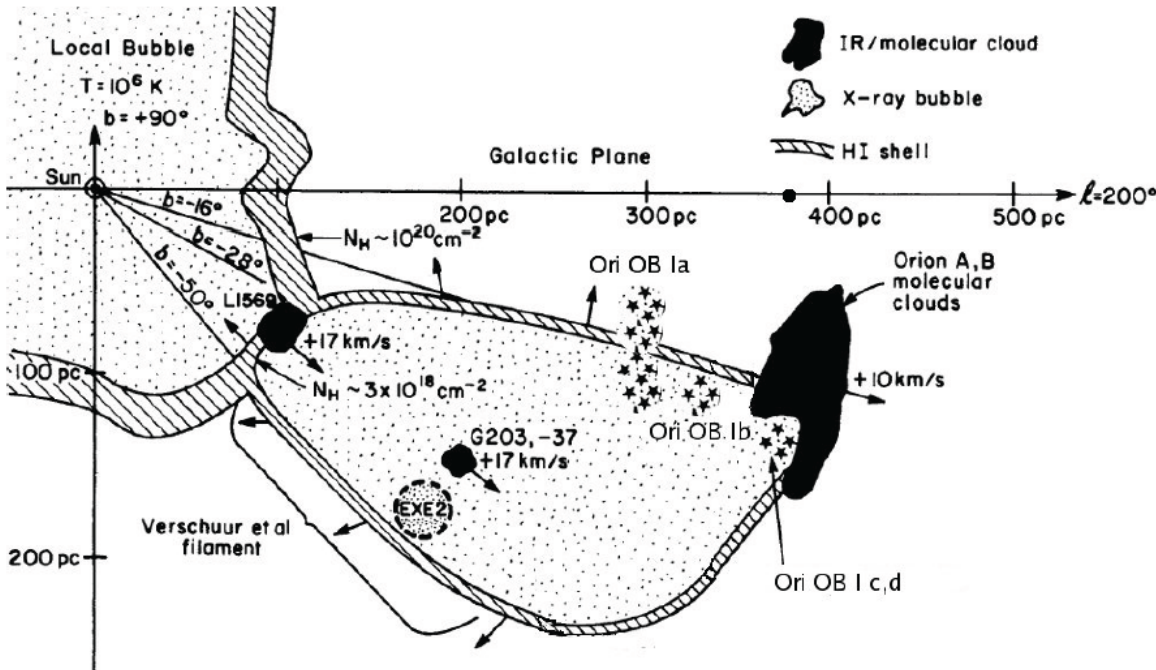}}
\centerline{
\includegraphics[width=0.65\textwidth]{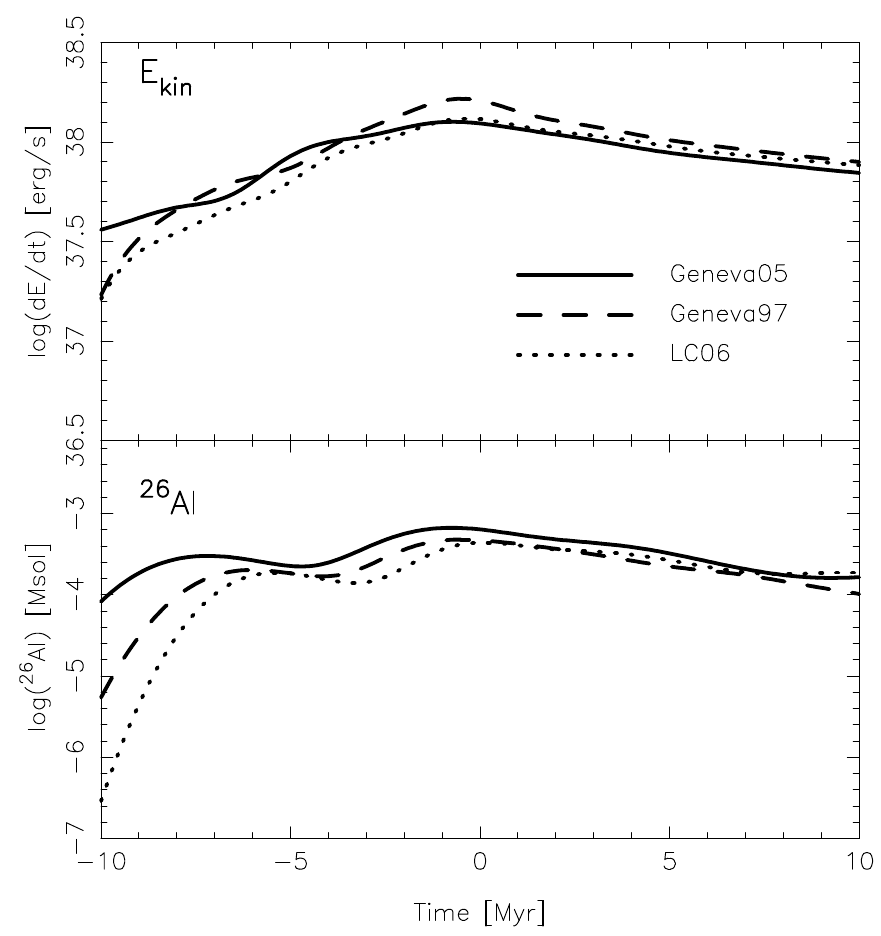}}
\caption{{\it (Top)} A sketch of a projection onto the Galactic plane of the region between us and the molecular clouds in Orion at about 450~pc distance, with the OB1 stellar association and its subgroups on the near side of the clouds, and the Eridanus cavity extending from the clouds towards the Sun \citep[from][]{Fierlinger:2012}. A scenario of \Al distribution from ejecta of the Orion OB1 association, blown into the Eridanus cavity, is indicated.
 {\it (Bottom)} The predicted time dependence of ejections of kinetic energy (above) and \Al (below), from population synthesis of the Orion OB1 association and its subgroups \citep{Voss:2010a}. }
\label{fig:26Al_eridanus-orion} 
\end{figure}   

Several star clusters at distances between 800 and 1500~pc were identified as candidate sources \citep{Pluschke:2002,Comeron:2002,Knodlseder:2000,Knodlseder:2002}.  
A population synthesis approach made use of the astronomical knowledge about these clusters, and derived predictions for the expected \Al emission, as well as the expected ionising luminosity from starlight, and kinetic energy injection from massive stars and supernovae that would lead to swept-up shells and cavities \citep{Cervino:2000,Voss:2009}.
Applying this to the Cygnus region \citep{Knodlseder:2002,Kretschmer:2003a,Martin:2009,Martin:2010b}, it became clear that age and metallicity of the major sources would drive the observed \Al emission (see Fig.~\ref{fig:al_cygnus}, right), and that Cyg OB2 was revealed to be a more-massive star cluster than had been thought before, as it is hidden behind molecular gas occulting starlight from its massive-star members.

\begin{figure}  
\centerline{\includegraphics[width=0.9\textwidth]{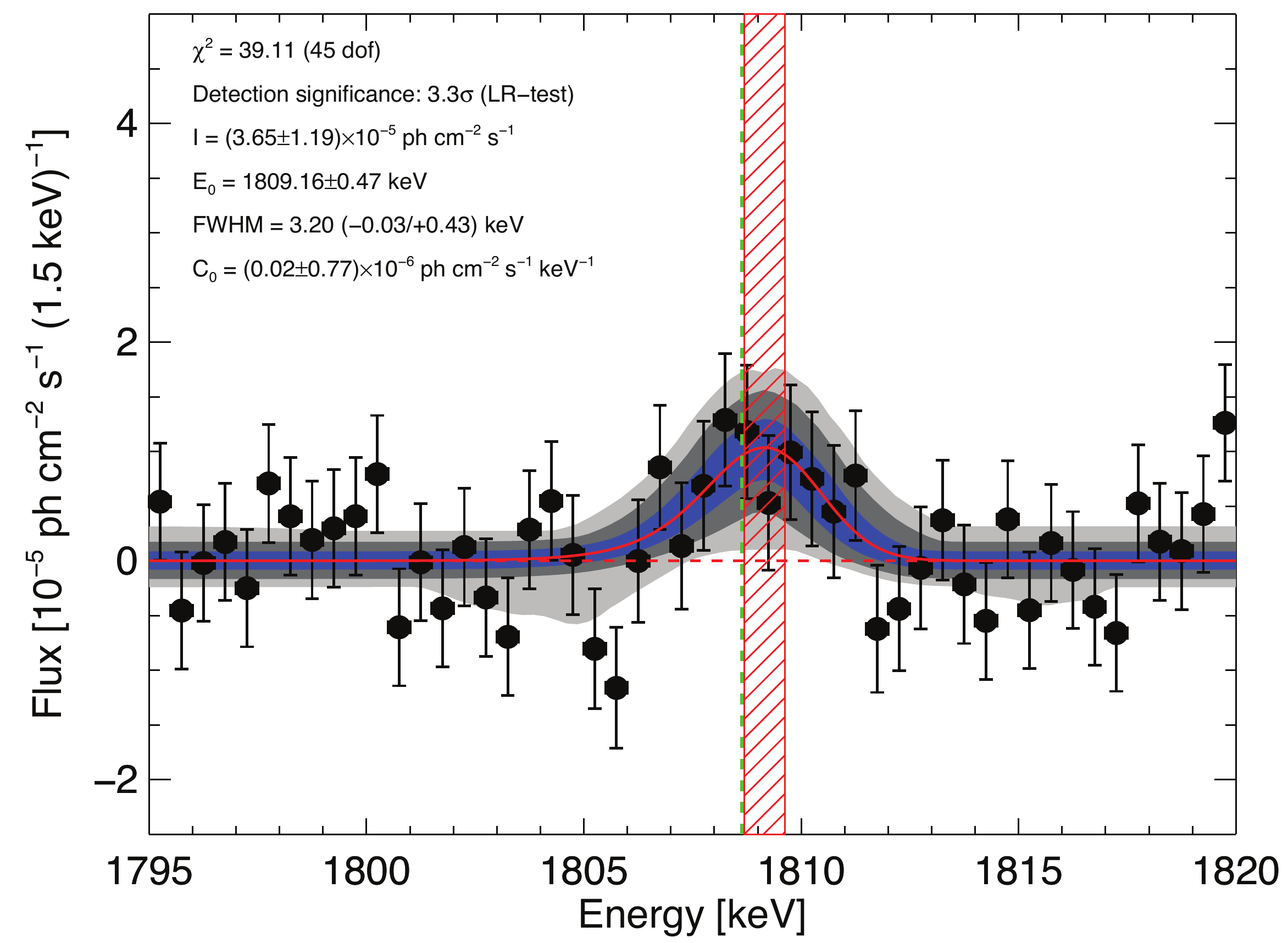}}\centerline{\includegraphics[width=0.9\textwidth]{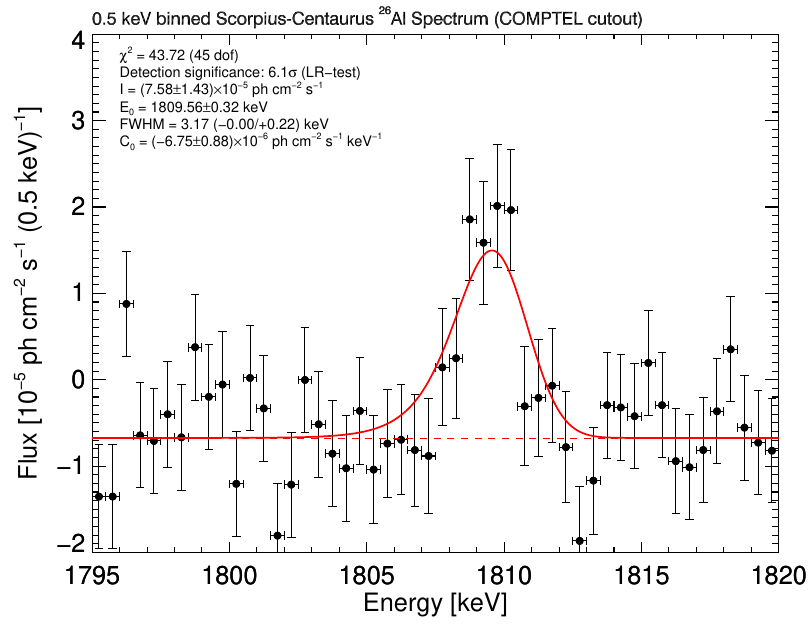}}
\caption{The \Al signal disentangled from the Orion region (above; \citep{Siegert:2017}), and from the Scorpius-Centaurus region  (below; \citep{Diehl:2010, Siegert:2017}). 
In the Orion-region spectrum, the hatched region indicates the uncertainty of the measured line centroid; it is slightly offset (1.1~$\sigma$) from the \Al laboratory energy of 1808.63 keV (vertical green-dashed line). }
\label{fig:26Al_sco-cen-orion} 
\end{figure}   

\begin{figure}  
\centerline{
\includegraphics[width=1.2\textwidth]{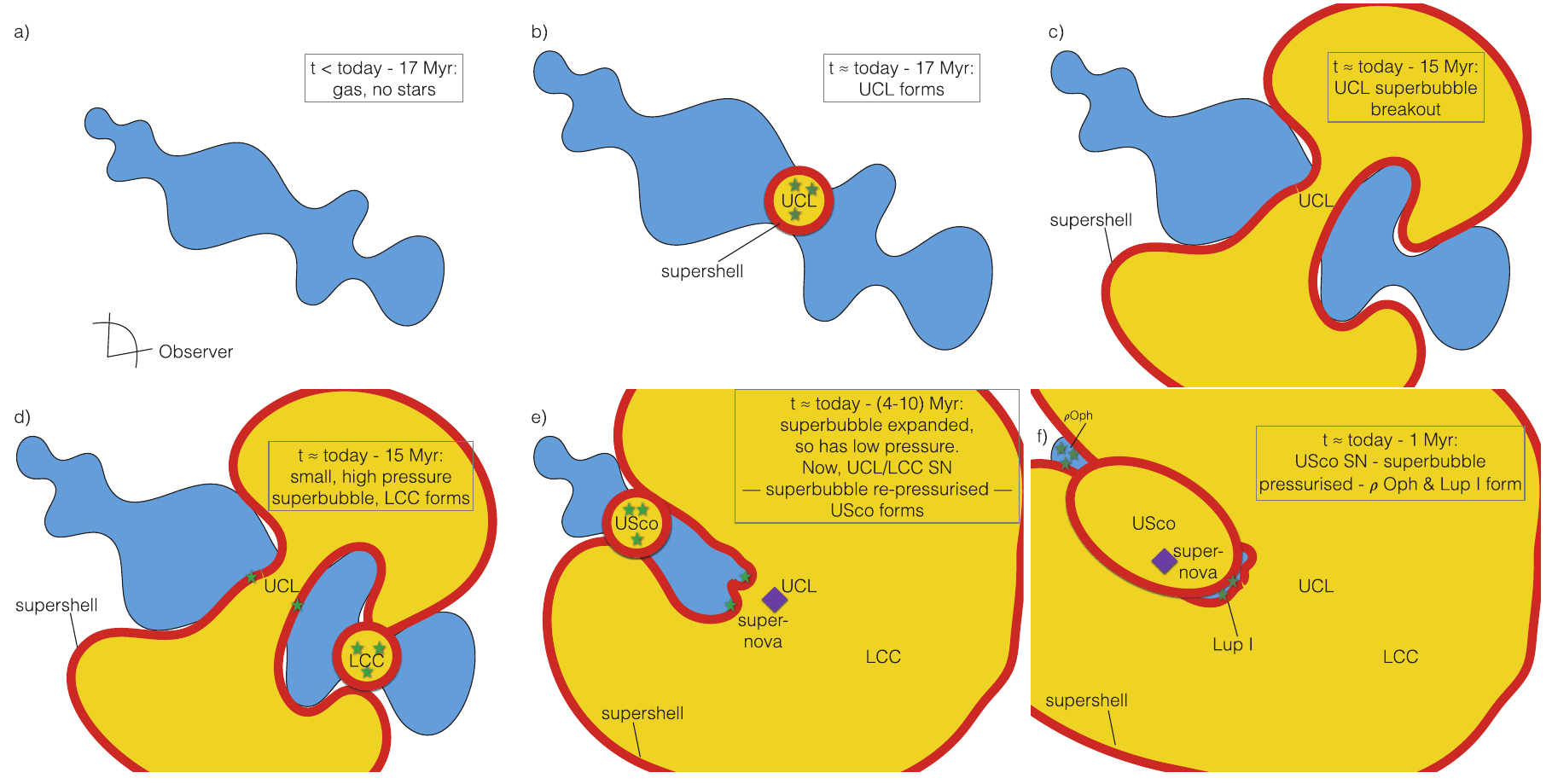}  }
\caption{An illustration of the scenario of successive erosion of star-forming molecular clouds by stellar feedback \citep{Krause:2018}. }
\label{fig:sco-cen-sketch} 
\end{figure}   

The Orion region is located towards the outer Galaxy, and holds one of the most-nearby massive star clusters, at a distance of $\approx$450~pc (see Fig.~\ref{fig:26Al_eridanus-orion}, top). The Orion OB1 cluster subgroups have some spread in ages, between 1.5 and 15 My \citep{Brown:1994}, which may indicate sequential, or even triggered, star formation within the former giant molecular cloud that is now visible as Orion A and Orion B. Interestingly, atomic HI data as well as X-ray data have identified a rather large cavity in the foreground of these sources \citep{Brown:1995}, extending towards the location of the Sun (see sketch in upper part of Fig.~\ref{fig:26Al_eridanus-orion}).  
From the different subgroup ages, population synthesis here predicts a rather flat time profile of \Al emission (see Fig.~\ref{fig:26Al_eridanus-orion}, bottom),  even rising in future \citep{Voss:2010a}.
\Al emission from Orion had first been reported from COMPTEL data \citep{Diehl:2003e}.
Cumulative INTEGRAL observations confirmed Orion as a source of \Al emission \citep{Siegert:2017a}, see Fig.~\ref{fig:26Al_sco-cen-orion}(right).

The Orion region with its foreground cavity provides an interesting nearby setting reminiscent to the superbubble scenario discussed above as an explanation of the large-scale Galactic Doppler shifts of the \Al line (see Fig.~\ref{fig:al_rotation-model}). In this case, a single cavity and cluster source set a unique geometry, as shown in Fig.~\ref{fig:26Al_eridanus-orion}. 
Simulations of the interstellar medium as it faces massive-star activity had demonstrated how one could expect to observe erosion of the parental molecular clouds and production of an outflow of hot gas \citep{Fierlinger:2016}.
It is interesting that the INTEGRAL measurement at high spectral resolution provides an indication for a blue shift of the \Al (see Fig.~\ref{fig:26Al_sco-cen-orion}, upper graph), as it is expected from the above scenario and this particular geometry.

 The Scorpius-Centaurus massive-star groups are even closer, at distances of 120--160~pc only \citep{Preibisch:2008}. Due to its proximity, the stars as well as the swept-up cavity interiors and walls are spread over rather large regions of the sky. This makes astronomical observations difficult for telescopes with modest fields of view, and thus led to debates about memberships of stars \citep{de-Geus:1989,de-Geus:1992,Rizzuto:2016}. 
 \Al emission from this region appeared only weakly indicated in the COMPTEL image.
 INTEGRAL measurements, however, allowed to discriminate \Al $\gamma$-ray emission from the Sco-Cen star clusters against the bright \Al emission from the Galactic plane,  see Fig.~\ref{fig:26Al_sco-cen-orion}(lower graph) \citep{Diehl:2010}. 
We conducted a multi-wavelength effort involving latest radio data from atomic hydrogen \citep{Winkel:2008}, as well as interpretations of Planck and Herschel data associated with star forming activity of the Scorpius-Centaurus region \citep{Gaczkowski:2017}, and simulations of the interstellar medium in 3D hydro. From this, we developed an understanding of how star formation may proceed successively in and around a parental molecular cloud \citep{Krause:2018}. The scenario is illustrated in Fig.~\ref{fig:sco-cen-sketch}, and shows that hot gas flowing out of a molecular cloud due to massive-star activity (as seen in Orion) would surround the cloud remains, and squash these through converging flows and shocks, thus triggering new star formation at different locations within the same cloud. 
 INTEGRAL measurements of \Al gamma rays have contributed significantly to establish this picture, which revises and modifies what was often discussed as \emph{triggered star formation} \citep{Preibisch:1999,Zinnecker:2007,Lee:2009}.


 \begin{figure}  
\centering 
\includegraphics[width=0.8\textwidth]{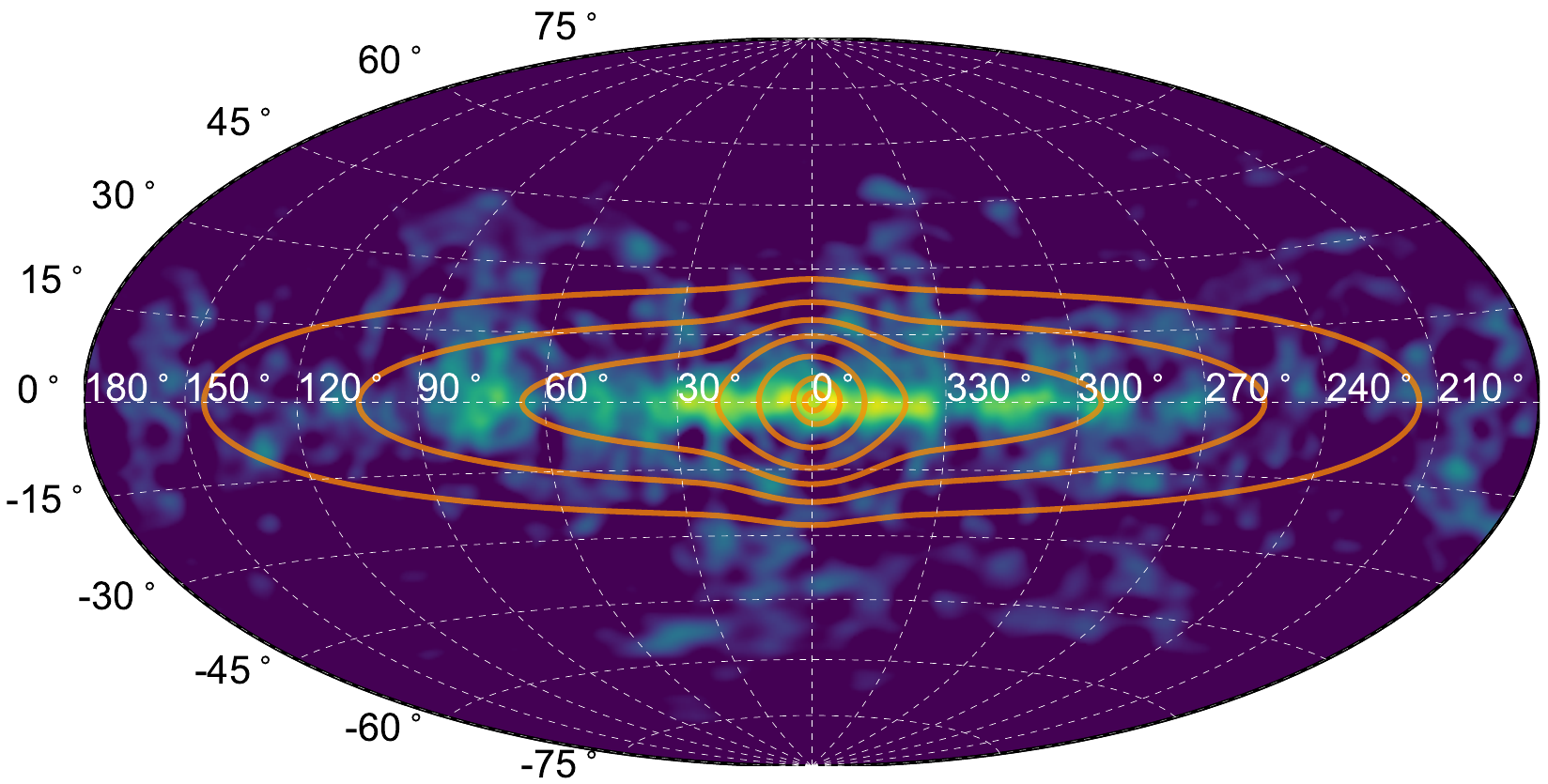}
\caption{The \Al image with contours overlaid from a best-fitting model reflecting the spatial morphology of the $\gamma$-ray emission from positron annihilation.}
\label{fig:26Al_511_maps} 
\end{figure}   

\subsection{Connections to INTEGRAL e$^+$ annihilation results} 
With INTEGRAL's energy range also including emission from positron annihilation near 500 keV and down to 20 keV, the positrons emitted from $\beta^+$~decays of radioactive by-products of nucleosynthesis (see Fig.~\ref{fig_26Al60Fe-decays}, left) can be associated with the diffuse $\gamma$-ray results of \Al \citep{Siegert:2017}.
Note that other isotopes such as $^{56}$Ni, $^{44}$Ti, and $^{22}$Na also contribute to such positrons from nucleosynthesis. As the radioactivity in this case would be more short-lived, such contributions may be less of a diffuse nature and could be localized to their source regions more than from \Al positrons. Therefore we focus our discussion on \Al connections. (See also the special article on positron annihilation $\gamma$ rays, also included in this Volume \citep{Churazov:2020}.)

From the \Al observations, one may use the decay scheme physics combined with the measured \Al $\gamma$-ray flux and an assumption about positron annihilation efficiency to estimate the sky brightness in 511~keV $\gamma$-rays from \Al positrons alone.  This led to a contribution estimate of order 10\% of \Al to the total observed Galactic $\gamma$-ray emission from positron annihilation\citep{Siegert:2017,Prantzos:2011}.
Adding contributions from other nucleosynthesis of $\beta^+$-decaying isotopes, it was shown that the combined nucleosynthesis contribution to positron annihilation $\gamma$-ray emission would be at odds with the Galactic longitude and latitude profiles of the positron annihilation $\gamma$-ray emission as observed by INTEGRAL. 
Fig.~\ref{fig:26Al_511_maps} illustrates a comparison of the $\gamma$-ray emission morphologies for \Al emission and for positron annihilation emission as seen by COMPTEL and INTEGRAL, respectively  \citep[from][]{Siegert:2017} \citep[see also][]{Siegert:2016,Knodlseder:2003}.

\section{The $^{60}$Fe isotope}  

$^{60}$Fe is a prominent neutron-rich isotope of Fe, which is radioactive with a decay time of 3.61~My \citep{Rugel:2009,Wallner:2015}.
The characteristic production reactions are neutron captures on abundant stable $^{54}$Fe to $^{56}$Fe nuclei; thus \Fe is produced by an s~process reaction flow, allowing for (only some!) intermediate $\beta$~decay as successive neutrons are captured by the nucleus. 
This keeps the reaction path  close to the region of stable isotopes, as is characteristic for the s~process. But $^{60}$Fe is two mass units away from this region, hence requires a suitably-long $\beta$-decay lifetime of intermediate nucleus $^{57}$Fe, and/or conditions that may involve higher neutron irradiation intensities, such as characteristic for an $i$- or  r~process.
Favourable production environments of \Fe are found in the helium-burning zones inside the more-massive stars above 6--8~\Msol, as 
here the  $^{22}$Ne($\alpha$,n)$^{25}$Mg reaction liberates neutrons \citep{Limongi:2006,Limongi:2006a}. 
Also in rare thermonuclear explosions, \Fe could be produced \citep{Woosley:1997}, although massive-stars have been considered more plausible \citep{Timmes:1995}.
Being sensitive to mixing and temperature structure in a massive star, the delicate production process of \Fe may tell us about the interiors of massive stars, through the yield in $^{60}$Fe. 
Different models have predicted yields that vary by as much as two orders of magnitude for stars at the higher mass end, i.e. beyond about 20-30~\Msol.

As discussed above, \Al ejection is attributed to massive stars mainly; therefore, INTEGRAL observations of \Al and of \Fe decay $\gamma$-rays may be unique probes of such massive-star physics, and the emission ratio as a diagnostic quantity would even eliminate uncertainties about source locations and distances in steady-state Galactic nucleosynthesis.
As such a steady state between current galactic production and decay of both these isotopes may be plausibly assumed, the determination of the ratio of their yields $r={Y_{60Fe}}/Y_{26Al} $ should be a very sensitive global diagnostic of the validity of massive-star nucleosynthesis models \citep[e.g.][and references therein]{Woosley:2007b}. 
According to models, \Fe is produced rather deep inside a massive star, in the Ne-O rich regions where He burning had been effective. This implies that  \Fe is expected to be ejected \emph{only} by the SN explosion and not by the stellar wind, contrary to the above case of \Al.
Current models predict a $\gamma$-ray flux ratio around 16\%  \citep{Timmes:1995,Woosley:2007b} or $\sim$18~($\pm$4)\% \citep{Limongi:2006a}.

Radioactive decay of \Fe (see Fig.~\ref{fig_26Al60Fe-decays}, bottom) emits $\gamma$-rays at energies of 1173 and 1332~keV, from cascade transitions towards the ground state within the final $^{60}$Ni daughter nucleus of the $^{60}$Fe$\rightarrow^{60}$Co$\rightarrow^{60}$Ni decay chain; a low-energy photon at 59~keV energy from a transition within the primary daughter $^{60}$Co occurs in 2\% of $^{60}$Fe-decays only, hence its brightness is below current instrument sensitivities. 

\Fe $\gamma$-rays have been hard to detect with current telescopes. 
The team of the Ramaty solar spectrometer mission RHESSI  first reported a marginal signal (2.6 $\sigma$ for the combined \Fe  1.173 and 1.332 MeV lines) \citep{Smith:2004}.
RHESSI also features Ge detectors, however with more modest spectral resolution as cosmic rays eroded charge collection, and no annealings were performed. Their Earth occultation, as the only available imaging information, suggested that the observed \Fe emission may arise from the inner Galaxy. 
Important was that the \Fe emission appeared only at the 10\%-level of \Al brightness. 

\subsection{INTEGRAL measurements and lessons} 

 \begin{figure}  
\centerline{
\includegraphics[width=0.8\textwidth]{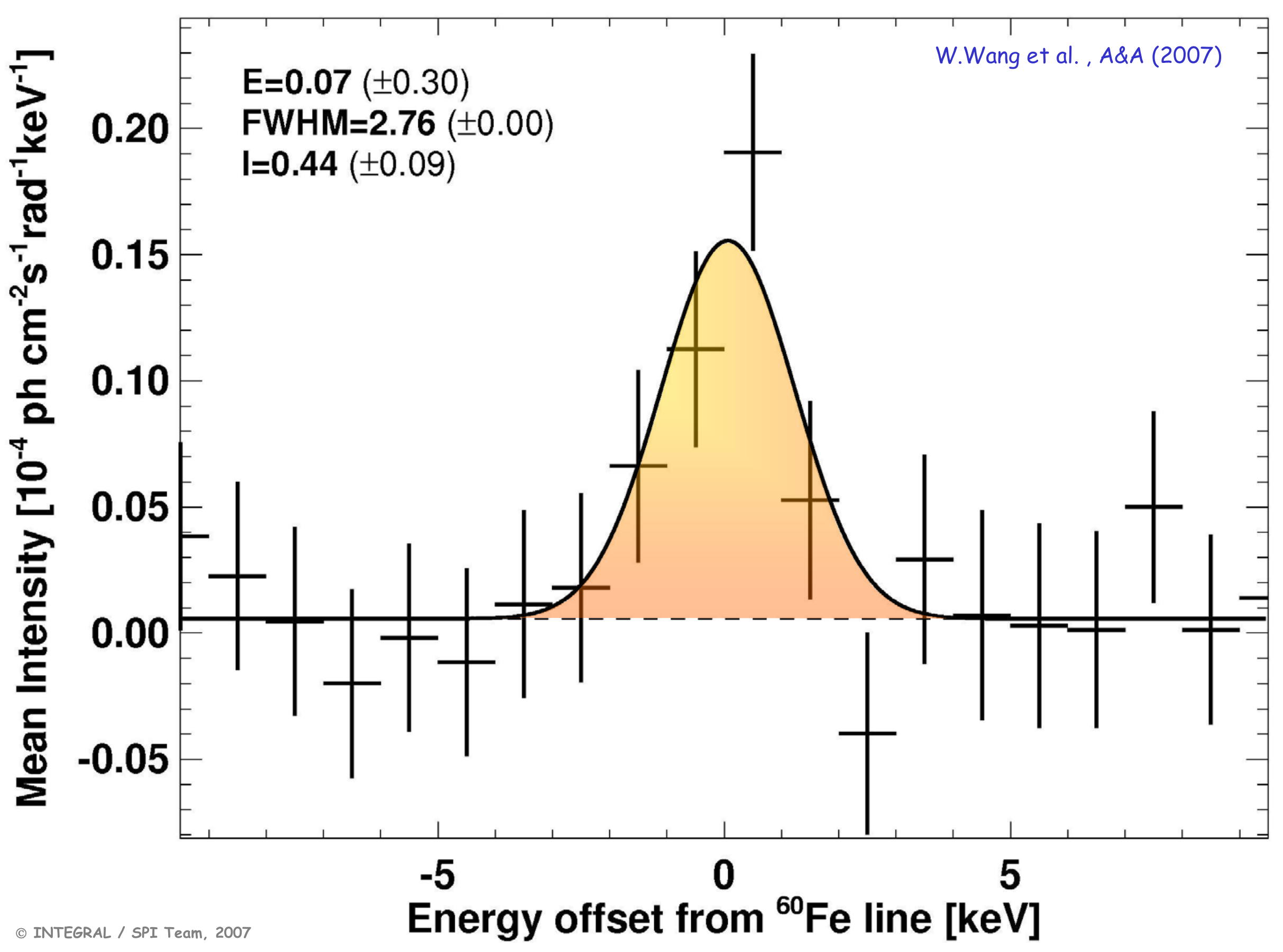}}\centerline{
\includegraphics[width=0.8\textwidth]{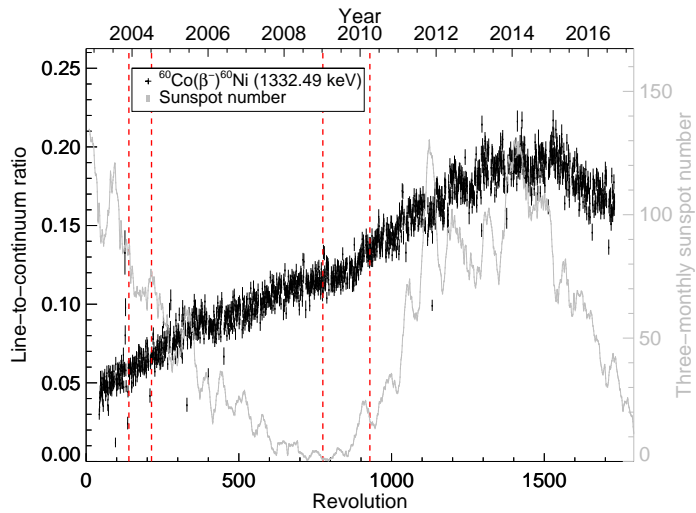}}
\caption{ {\it (Top:)} The \Fe signal (both $\gamma$-ray lines superimposed at their laboratory energies of 1173 and 1332~keV) as observed with INTEGRAL/SPI from the Galaxy \citep{Wang:2007}
 {\it (Bottom:)} The intensity of background line emission from $^{60}$Co, here scaled with the varying continuum intensity of background. This radioactivity is built up from cosmic-ray bombardment of INTEGRAL in orbit, hence grows in relation to continuum background. The \Fe result shown on the left derives from data in orbits 17-359, i.e. the early part of the mission. Note that instrumental background varies somewhat, in anti-correlation with solar activity, which is shown as the grey line.}
\label{fig:60Fe_SPI} 
\end{figure}   

INTEGRAL observations in the early mission years could reveal a weak signal from \Fe in the characteristic $\gamma$-ray lines at 1173 and 1332 keV: 
Fig.~\ref{fig:60Fe_SPI} (top graph) shows the signal in both lines superimposed \citep{Wang:2007}, which were derived after about 3 mission years of data.  

\Fe detectability with INTEGRAL suffers from a specific problem: The isotope $^{60}$Co apparently is built up within the INTEGRAL spacecraft due to irradiation of the spacecraft materials with cosmic rays. With a radioactive lifetime of 7.6 years, build-up and decay should reach a steady state within a few lifetimes, and a quasi-linear growth is seen during the first $\approx$10 mission years in Fig.~\ref{fig:60Fe_SPI} (lower graph) \citep{Diehl:2018}. 
From the data of the first mission years, a \Fe to \Al brightness ratio of 14~($\pm$6)\% was reported  \citep{Wang:2007}.

Although we expect the sources of \Al and \Fe to be the same category of massive stars, there are, however, some important differences:
First, the radioactive lifetimes are different, \Fe being 3.46 times longer-lived; this could make \Fe emission more diffuse, from the same source, if propagation is the same.
Second, ejection of \Fe only occurs at the time of the core-collapse supernova, while \Al ejection may start already in the strong-wind phase of a massive star's evolution. This implies that the interstellar medium has been pre-shaped by winds before \Fe ejection, and thus the channeling of \Al ejecta by surrounding gas would be more restrictive than for $^{60}$Fe.
These plausible cases can best be investigated in 3D-hydrodynamic simulations. Such have been made recently at large-enough (galactic) scale \citep{Fujimoto:2018,Rodgers-Lee:2019}.
In principle, those simulations confirm the above simple arguments.
However, from the limitations of required and  different assumptions about the basic structure within a galaxy, e.g. spiral arms and gravitational fields from stars, gas, and dark matter, no quantitative conclusions can be derived yet from the \Fe measurements.
Hence, we must address the unknown morphology of \Fe $\gamma$-ray line emission through different ways. However, being weak compared to \Al emission, construction of an image of \Fe emission is not feasible from INTEGRAL data.
In particular, a challenge is the important test whether the spatial distribution of \Fe emission is identical or not to that of \Al.

 \begin{figure}  
\centerline{
\includegraphics[width=0.6\textwidth]{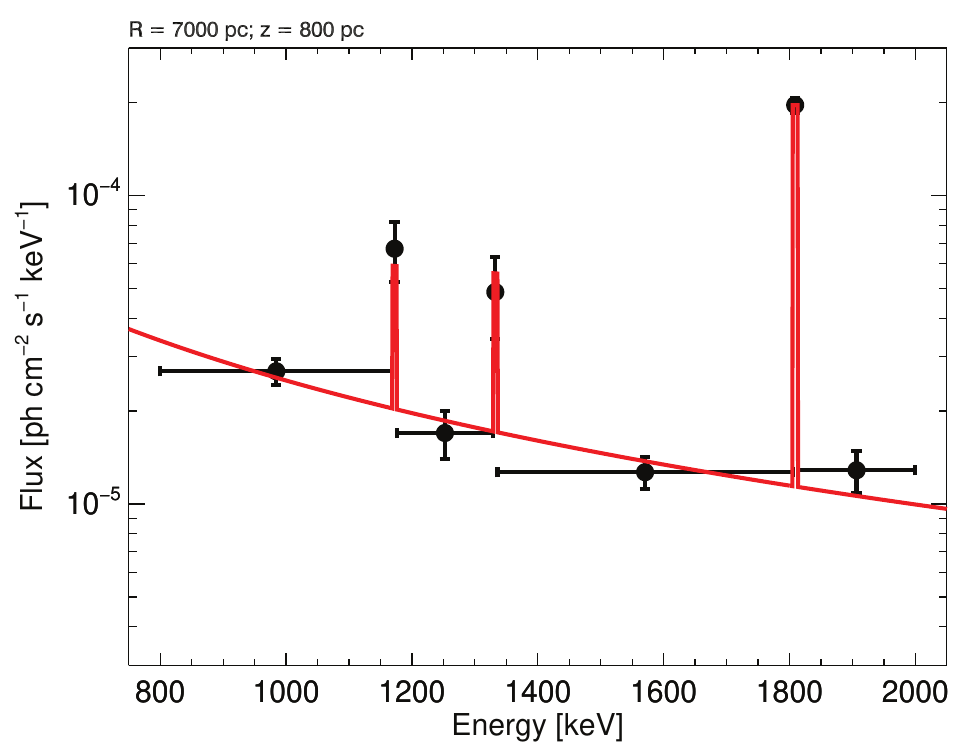}
\includegraphics[width=0.6\textwidth]{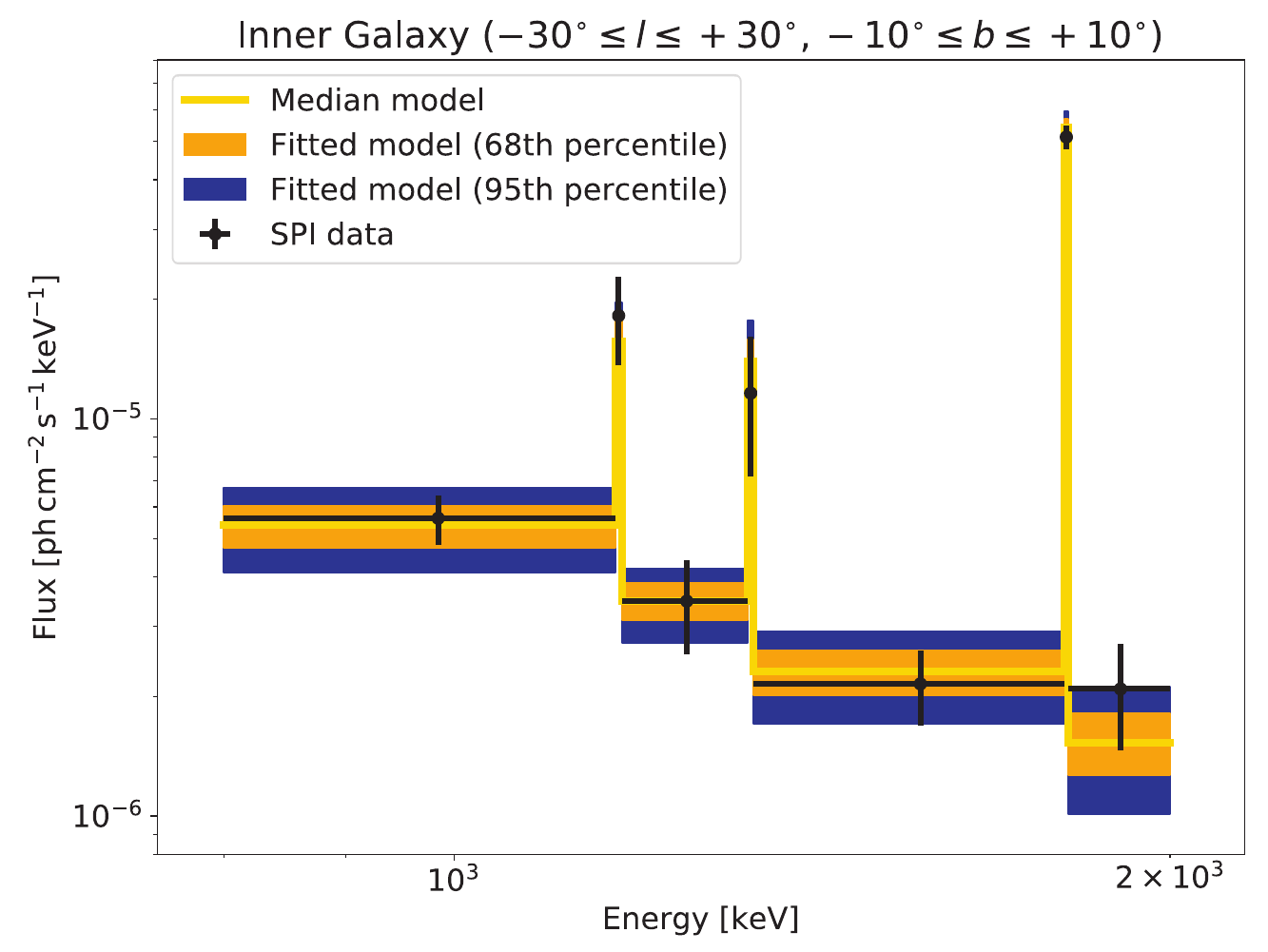}}
\caption{ {\it (Left:)} The spectrum of diffuse $\gamma$-ray emission in the 0.8--2.0~MeV range, with continuum bands and lines from \Fe at 1173 and 1332~keV and from \Al at 1809 keV. Modelling the sky as a diffuse exponential disk {\it (left)}) or by the COMPTEL \Al image {\it (right})  leads to different values in detail, illustrating uncertainties from the emission morphology. \citep[From][]{Wang:2020}}
\label{fig:60Fe26AlCont_SPI} 
\end{figure}   

 \begin{figure}  
\centerline{
\includegraphics[width=0.8\textwidth]{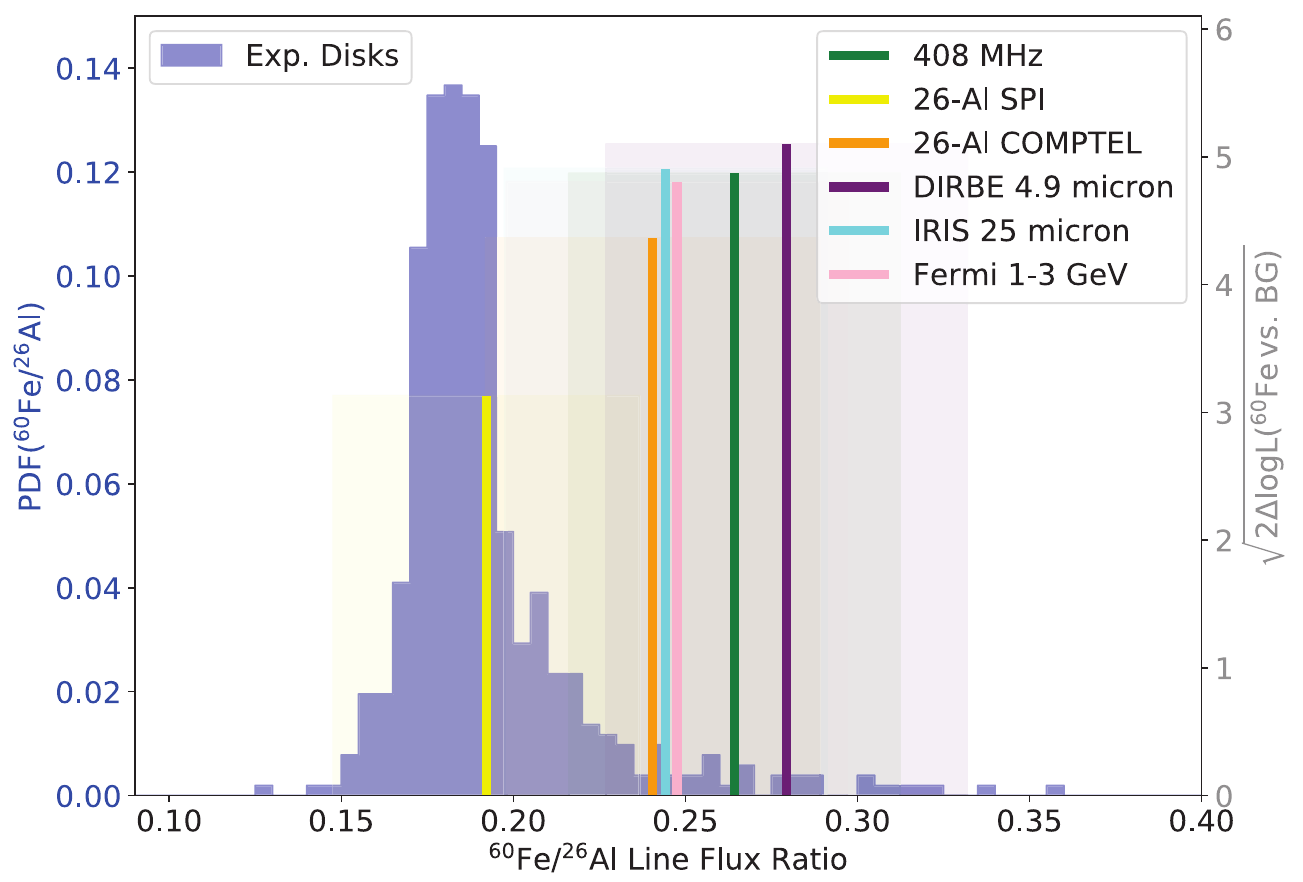}}
\caption{  The \Fe/\Al ratio derived from SPI observations depends on the adopted emission morphology (see Fig.~\ref{fig:60Fe26AlCont_SPI}). From a range of exponential-disk models with different parameters (dark-blue histogram) through models from specific source tracers (color bars), a plausibly-constrained range of 0.2--0.4 emerges. \citep[From][]{Wang:2020}}
\label{fig:60Fe26Al-ratio_SPI} 
\end{figure}   

A new attempt has been made, using all available accumulated INTEGRAL data, and exploring results for a wide range of potential morphologies of \Fe emission  \citep{Wang:2020}. Here, within a broad band of data from 0.8 to 2 MeV, several continuum bands and the lines from \Fe and from \Al were fitted consistently to the same adopted morphology.
Exponential-disk morphologies could be varied through their scale radius and scale height parameters within a broad plausible range. For comparison, astronomical tracers such as the \Al image from COMPTEL or SPI, dust emission, and a range of other, more-or-less far-fetched tracers were used. 
Fig.~\ref{fig:60Fe26AlCont_SPI} shows an example of spectra obtained for a typical exponential disk model that rather closely resembles \Al (but maybe not $^{60}$Fe) emission (left), and for the COMPTEL \Al image within the inner Galaxy (right).
Fig.~\ref{fig:60Fe26Al-ratio_SPI} shows the resulting $\gamma$-ray line intensity ratios \Fe/\Al for such a broad range of potential morphologies. 
This analysis found that for \Fe data alone, morphologies akin a concentrated point source do not fit the data well, being inferior in quality by a significant amount compared to diffuse and extended morphologies. This alone confirms the essentially-diffuse nature of \Fe emission. But no further imaging constraints are feasible from INTEGRAL data so far.
On the other hand, the   \Fe/\Al ratio varies  ($\pm$4\%) around 18.4\% for all exponential-disk variants. Other plausible tracers apparently obtain significantly higher ratio values such as 0.28 for the dust emission map. 
From this analysis, a range for the   \Fe/\Al ratio of 0.2 -- 0.4 has been derived, in account of these systematic uncertainties on \Fe emission morphology, for the Galactic diffuse line emission  \citep{Wang:2020}. 
 
\section*{Summary}
INTEGRAL with its $\gamma$-ray spectrometer SPI has measured diffuse nucleosynthesis line emission from the \Al and \Fe isotopes, and (less directly) from positron annihilation emission with its 511~keV line.
\Al measurements have established the current production within the extended entire Galaxy, most-plausibly ejected  from massive stars and their supernovae.
The investigations of specific massive star clusters which are within distances of order one kpc have allowed specific tests of the massive-star origin hypothesis. Population synthesis model comparisons and multi-frequency studies have been carried through, analysing other implications of massive-star activity such as large cavities recognised through emission of hot gas in their interiors and 21cm H1 emission in their swept-up shells.
On the galactic scale, Doppler shifts and broadenings of the \Al $\gamma$-ray line have been compared with kinematic data from other objects. This has led to the insight that production of large cavities and superbubbles are a typical characteristic of interstellar medium into which \Al is injected from its sources.
\Fe $\gamma$-ray emission has been clearly detected, and is found significantly weaker in intensity than \Al emission. This, and the radioactive build-up of a particular background in $^{60}$Co emitting the same lines, makes astrophysical analysis for \Fe much more challenging. 
\Fe most probably also is diffuse in nature within our Galaxy, rather than from a few bright individual sources. 
The intensity ratio found from INTEGRAL of $^{60}$Fe/\Al is constrained to lie in the range 0.2--0.4. This appears consistent with constraints derived from solar-system probes such as cosmic rays \citep{Kachelries:2015} and deposits on Earth and Moon \citep{Wallner:2016}.  

In summary, INTEGRAL results from 15 years of data have reached a precision that enables detailed astrophysical studies for \Al, and \Fe data are sufficient for coarse tests of astrophysical models. 
Future mission years will enhance the statistical precision, but even more-importantly the recognition of systematics from the large instrumental background. This should increase the overall sensitivity for diffuse nucleosynthesis approximately linearly with time, stimulating hope for more astrophysical results in this field before its mission end in 2029.
Other diffuse emission in our Galaxy may, e.g., be expected from $^{22}$Na radioactivity as produced by novae, or from long-lived r-process radioactivity produced in neutron star mergers.

\section*{Acknowledgements}
  We are grateful for constructive comments of the referee, which helped the quality of this review paper. The INTEGRAL/SPI project
  has been completed under the responsibility and leadership of CNES;
  we are grateful to ASI, CEA, CNES, DLR (No. 50 OG 1101, 1601, and 1901, and No. 50 OR 0901, 1206, 1611, and 1316), ESA, INTA, NASA and OSTC for
  support of this ESA space science mission.
  This research was also supported by the DFG cluster of excellence \emph{Origin and Structure of the Universe}, and by the ChETEC  Action (CA16117) of COST (European Cooperation in Science and Technology) .

\vfil\eject



\newcommand{\actaa}{Acta Astron. }%
\newcommand{\araa}{Ann.Rev.Astron.\&Astroph. }%
\newcommand{\apj}{Astroph.J. }%
\newcommand{\apjl}{Astroph.J.Lett. }%
\newcommand{\apjs}{Astroph.J.Supp. }%
\newcommand{\ao}{Appl.~Opt. }%
\newcommand{\apss}{Astroph.J.\&Sp.Sci. }%
\newcommand{\aap}{Astron.\&Astroph. }%
\newcommand{\aapr}{Astron.\&Astroph.~Rev. }%
\newcommand{\aaps}{Astron.\&Astroph.~Suppl. }%
\newcommand{\aj}{Astron.Journ. }%
\newcommand{\azh}{AZh }%
\newcommand{\memras}{MmRAS }%
\newcommand{\mnras}{Mon.Not.Royal~Astr.~Soc. }%
\newcommand{\na}{New Astron. }%
\newcommand{\nar}{New Astron. Rev. }%
\newcommand{\pra}{Phys.~Rev.~A }%
\newcommand{\prb}{Phys.~Rev.~B }%
\newcommand{\prc}{Phys.~Rev.~C }%
\newcommand{\prd}{Phys.~Rev.~D }%
\newcommand{\pre}{Phys.~Rev.~E }%
\newcommand{\prl}{Phys.~Rev.~Lett. }%
\newcommand{\pasa}{PASA }%
\newcommand{\pasp}{Proc.Astr.Soc.Pac. }%
\newcommand{\pasj}{Proc.Astr.Soc.Jap. }%
\newcommand{\rpp}{Rep.Prog.Phys. }%
\newcommand{\skytel}{Sky\&Tel. }%
\newcommand{\solphys}{Sol.~Phys. }%
\newcommand{\sovast}{Soviet~Ast. }%
\newcommand{\ssr}{Space~Sci.~Rev. }%
\newcommand{\nat}{Nature }%
\newcommand{\iaucirc}{IAU~Circ. }%
\newcommand{\aplett}{Astrophys.~Lett. }%
\newcommand{\apspr}{Astrophys.~Space~Phys.~Res. }%
\newcommand{\nphysa}{Nucl.~Phys.~A }%
\newcommand{\physrep}{Phys.~Rep. }%
\newcommand{\procspie}{Proc.~SPIE }%
         



\begin{thebibliography}{100}
\expandafter\ifx\csname url\endcsname\relax
  \def\url#1{\texttt{#1}}\fi
\expandafter\ifx\csname urlprefix\endcsname\relax\def\urlprefix{URL }\fi
\expandafter\ifx\csname href\endcsname\relax
  \def\href#1#2{#2} \def\path#1{#1}\fi

\bibitem{Clayton:1993}
D.~D. {Clayton}, D.~H. {Hartmann}, M.~D. {Leising}, {On Al-26 and other
  short-lived interstellar radioactivity}, \apjl 415 (1993) L25--L29.
\newblock \href {http://dx.doi.org/10.1086/187024} {\path{doi:10.1086/187024}}.

\bibitem{Schaller:1992}
G.~{Schaller}, D.~{Schaerer}, G.~{Meynet}, A.~{Maeder}, {New grids of stellar
  models from 0.8 to 120 solar masses at Z = 0.020 and Z = 0.001}, \aaps 96
  (1992) 269--331.

\bibitem{Signore:1993}
M.~{Signore}, C.~{Dupraz}, {Massive stars as Galactic producers of Al-26},
  \aaps 97 (1993) 141--144.

\bibitem{Meynet:1997}
G.~{Meynet}, M.~{Arnould}, N.~{Prantzos}, G.~{Paulus}, {Contribution of
  Wolf-Rayet stars to the synthesis of $^{26}$Al. I. The {$\gamma$}-ray
  connection.}, \aap 320 (1997) 460--468.

\bibitem{Bazan:1993}
G.~{Bazan}, L.~{Brown}, D.~{Clayton}, J.~{Truran}, {26AL Production in AGB
  Stars and 1.809 MeV Line Emission in the Galaxy}, Revista Mexicana de
  Astronomia y Astrofisica 27 (1993) 87--+.

\bibitem{Forestini:1997}
M.~{Forestini}, C.~{Charbonnel}, {Nucleosynthesis of light elements inside
  thermally pulsing AGB stars: I. The case of intermediate-mass stars}, \aaps
  123 (1997) 241--272.
\newblock \href {http://dx.doi.org/10.1051/aas:1997348}
  {\path{doi:10.1051/aas:1997348}}.

\bibitem{Karakas:2007}
A.~{Karakas}, J.~C. {Lattanzio}, {Stellar Models and Yields of Asymptotic Giant
  Branch Stars}, Publications of the Astronomical Society of Australia 24
  (2007) 103--117.
\newblock \href {http://arxiv.org/abs/0708.4385} {\path{arXiv:0708.4385}},
  \href {http://dx.doi.org/10.1071/AS07021} {\path{doi:10.1071/AS07021}}.

\bibitem{Timmes:1995}
F.~X. {Timmes}, S.~E. {Woosley}, D.~H. {Hartmann}, R.~D. {Hoffman}, T.~A.
  {Weaver}, F.~{Matteucci}, {26Al and 60Fe from Supernova Explosions}, \apj 449
  (1995) 204--+.
\newblock \href {http://arxiv.org/abs/arXiv:astro-ph/9503120}
  {\path{arXiv:arXiv:astro-ph/9503120}}, \href
  {http://dx.doi.org/10.1086/176046} {\path{doi:10.1086/176046}}.

\bibitem{Prantzos:1996a}
N.~{Prantzos}, R.~{Diehl}, {Radioactive 26Al in the galaxy: observations versus
  theory}, \physrep 267 (1996) 1--69.
\newblock \href {http://dx.doi.org/10.1016/0370-1573(95)00055-0}
  {\path{doi:10.1016/0370-1573(95)00055-0}}.

\bibitem{Jose:1998}
J.~{Jose}, M.~{Hernanz}, {Nucleosynthesis in Classical Novae: CO versus ONe
  White Dwarfs}, \apj 494 (1998) 680--+.
\newblock \href {http://arxiv.org/abs/arXiv:astro-ph/9709153}
  {\path{arXiv:arXiv:astro-ph/9709153}}, \href
  {http://dx.doi.org/10.1086/305244} {\path{doi:10.1086/305244}}.

\bibitem{Karakas:2014}
A.~I. {Karakas}, J.~C. {Lattanzio}, {The Dawes Review 2: Nucleosynthesis and
  Stellar Yields of Low- and Intermediate-Mass Single Stars}, \pasa 31 (2014)
  e030.
\newblock \href {http://arxiv.org/abs/1405.0062} {\path{arXiv:1405.0062}},
  \href {http://dx.doi.org/10.1017/pasa.2014.21}
  {\path{doi:10.1017/pasa.2014.21}}.

\bibitem{Clayton:2004}
D.~D. {Clayton}, L.~R. {Nittler}, {Astrophysics with Presolar Stardust}, \araa
  42 (2004) 39--78.
\newblock \href {http://dx.doi.org/10.1146/annurev.astro.42.053102.134022}
  {\path{doi:10.1146/annurev.astro.42.053102.134022}}.

\bibitem{Mahoney:1984}
W.~A. {Mahoney}, J.~C. {Ling}, W.~A. {Wheaton}, A.~S. {Jacobson}, {HEAO 3
  discovery of Al-26 in the interstellar medium}, \apj 286 (1984) 578--585.
\newblock \href {http://dx.doi.org/10.1086/162632} {\path{doi:10.1086/162632}}.

\bibitem{Pluschke:2001c}
S.~{Pl{\"u}schke}, R.~{Diehl}, V.~{Sch{\"o}nfelder}, H.~{Bloemen},
  W.~{Hermsen}, K.~{Bennett}, C.~{Winkler}, M.~{McConnell}, J.~{Ryan},
  U.~{Oberlack}, J.~{Kn{\"o}dlseder}, {The COMPTEL 1.809 MeV survey}, in:
  A.~{Gimenez}, V.~{Reglero}, C.~{Winkler} (Eds.), Exploring the Gamma-Ray
  Universe, Vol. 459 of ESA Special Publication, 2001, pp. 55--58.

\bibitem{Bouchet:2015}
L.~{Bouchet}, E.~{Jourdain}, J.-P. {Roques}, {The Galactic $^{26}$Al Emission
  Map as Revealed by INTEGRAL SPI}, \apj 801 (2015) 142.
\newblock \href {http://arxiv.org/abs/1501.05247} {\path{arXiv:1501.05247}},
  \href {http://dx.doi.org/10.1088/0004-637X/801/2/142}
  {\path{doi:10.1088/0004-637X/801/2/142}}.

\bibitem{Diehl:1995b}
R.~{Diehl}, C.~{Dupraz}, K.~{Bennett}, H.~{Bloemen}, W.~{Hermsen},
  J.~{Knoedlseder}, G.~{Lichti}, D.~{Morris}, J.~{Ryan}, V.~{Schoenfelder},
  H.~{Steinle}, A.~{Strong}, B.~{Swanenburg}, M.~{Varendorff}, C.~{Winkler},
  {COMPTEL observations of Galactic $^{26}$Al emission.}, \aap 298 (1995)
  445--+.

\bibitem{Oberlack:1996}
U.~{Oberlack}, K.~{Bennett}, H.~{Bloemen}, R.~{Diehl}, C.~{Dupraz},
  W.~{Hermsen}, J.~{Knoedlseder}, D.~{Morris}, V.~{Schoenfelder}, A.~{Strong},
  C.~{Winkler}, {The COMPTEL 1.809MeV all-sky image.}, \aaps 120 (1996) C311.

\bibitem{Knodlseder:1999}
J.~{Kn{\"o}dlseder}, D.~{Dixon}, K.~{Bennett}, H.~{Bloemen}, R.~{Diehl},
  W.~{Hermsen}, U.~{Oberlack}, J.~{Ryan}, V.~{Sch{\"o}nfelder}, P.~{von
  Ballmoos}, {Image reconstruction of COMPTEL 1.8 MeV (26) AL line data}, \aap
  345 (1999) 813--825.
\newblock \href {http://arxiv.org/abs/arXiv:astro-ph/9903172}
  {\path{arXiv:arXiv:astro-ph/9903172}}.

\bibitem{Knodlseder:1999c}
J.~{Kn{\"o}dlseder}, D.~{Dixon}, R.~{Diehl}, U.~{Oberlack}, P.~{von Ballmoos},
  H.~{Bloemen}, W.~{Hermsen}, A.~J. {Iyudin}, J.~{Ryan}, V.~{Sch{\"o}nfelder},
  {A New Look at Image Reconstruction of COMPTEL 1.8 Mev Data}, Astrophysical
  Letters Communications 38 (1999) 395--+.
\newblock \href {http://arxiv.org/abs/arXiv:astro-ph/9902283}
  {\path{arXiv:arXiv:astro-ph/9902283}}.

\bibitem{Strong:1999a}
A.~W. {Strong}, H.~{Bloemen}, R.~{Diehl}, W.~{Hermsen}, V.~{Sch{\"o}nfelder},
  {COMPTEL Skymapping: a New Approach Using Parallel Computing}, Astrophysical
  Letters Communications 39 (1999) 209--+.
\newblock \href {http://arxiv.org/abs/arXiv:astro-ph/9811211}
  {\path{arXiv:arXiv:astro-ph/9811211}}.

\bibitem{Lingenfelter:1978}
R.~E. {Lingenfelter}, R.~{Ramaty}, {Gamma-ray lines - A new window to the
  Universe}, Physics Today 31 (1978) 40--47.

\bibitem{Prantzos:1993}
N.~{Prantzos}, {Radioactive Al-26 from massive stars - Production and
  distribution in the Galaxy}, \apjl 405 (1993) L55--L58.
\newblock \href {http://dx.doi.org/10.1086/186764} {\path{doi:10.1086/186764}}.

\bibitem{Lada:2003}
C.~J. {Lada}, E.~A. {Lada}, {Embedded Clusters in Molecular Clouds}, \araa 41
  (2003) 57--115.
\newblock \href {http://arxiv.org/abs/arXiv:astro-ph/0301540}
  {\path{arXiv:arXiv:astro-ph/0301540}}, \href
  {http://dx.doi.org/10.1146/annurev.astro.41.011802.094844}
  {\path{doi:10.1146/annurev.astro.41.011802.094844}}.

\bibitem{Lada:2005}
C.~J. {Lada}, {Star Formation in the Galaxy: An Observational Overview},
  Progress of Theoretical Physics Supplement 158 (2005) 1--23.
\newblock \href {http://dx.doi.org/10.1143/PTPS.158.1}
  {\path{doi:10.1143/PTPS.158.1}}.

\bibitem{Dobbs:2014a}
C.~{Dobbs}, J.~{Baba}, {Dawes Review 4: Spiral Structures in Disc Galaxies},
  \pasa 31 (2014) e035.
\newblock \href {http://arxiv.org/abs/1407.5062} {\path{arXiv:1407.5062}},
  \href {http://dx.doi.org/10.1017/pasa.2014.31}
  {\path{doi:10.1017/pasa.2014.31}}.

\bibitem{Reid:2014}
M.~J. {Reid}, K.~M. {Menten}, A.~{Brunthaler}, X.~W. {Zheng}, T.~M. {Dame},
  Y.~{Xu}, Y.~{Wu}, B.~{Zhang}, A.~{Sanna}, M.~{Sato}, K.~{Hachisuka}, Y.~K.
  {Choi}, K.~{Immer}, L.~{Moscadelli}, K.~L.~J. {Rygl}, A.~{Bartkiewicz},
  {Trigonometric Parallaxes of High Mass Star Forming Regions: The Structure
  and Kinematics of the Milky Way}, \apj 783~(2) (2014) 130.
\newblock \href {http://arxiv.org/abs/1401.5377} {\path{arXiv:1401.5377}},
  \href {http://dx.doi.org/10.1088/0004-637X/783/2/130}
  {\path{doi:10.1088/0004-637X/783/2/130}}.

\bibitem{del-Rio:1996}
E.~{del Rio}, P.~{von Ballmoos}, K.~{Bennett}, H.~{Bloemen}, R.~{Diehl},
  W.~{Hermsen}, J.~{Knoedlseder}, U.~{Oberlack}, J.~{Ryan}, V.~{Schoenfelder},
  C.~{Winkler}, {1.8 MeV line emission from the Cygnus region.}, \aap 315
  (1996) 237--242.

\bibitem{Knodlseder:2000}
J.~{Kn{\"o}dlseder}, {Cygnus OB2 - a young globular cluster in the Milky Way},
  \aap 360 (2000) 539--548.
\newblock \href {http://arxiv.org/abs/arXiv:astro-ph/0007442}
  {\path{arXiv:arXiv:astro-ph/0007442}}.

\bibitem{Pluschke:2000}
S.~{Pl{\"u}schke}, K.~{Kretschmer}, R.~{Diehl}, D.~H. {Hartmann},
  U.~{Oberlack}, {OB Associations as Sources of Galactic Radioactivity}, in:
  R.~E. {Schielicke} (Ed.), Astronomische Gesellschaft Meeting Abstracts,
  Vol.~16 of Astronomische Gesellschaft Meeting Abstracts, 2000, pp. 77--+.

\bibitem{Comeron:1994a}
F.~{Comeron}, J.~{Torra}, {The Formation of Stellar Systems by Gravitational
  Instability in the Cygnus Superbubble}, \apj 423 (1994) 652--+.
\newblock \href {http://dx.doi.org/10.1086/173843} {\path{doi:10.1086/173843}}.

\bibitem{Pluschke:2001a}
S.~{Pl{\"u}schke}, K.~{Kretschmer}, R.~{Diehl}, D.~H. {Hartmann}, U.~G.
  {Oberlack}, {The Cygnus region: $^{26}$Al from OB associations}, in:
  A.~{Gimenez}, V.~{Reglero}, C.~{Winkler} (Eds.), Exploring the Gamma-Ray
  Universe, Vol. 459 of ESA Special Publication, 2001, pp. 91--95.

\bibitem{Pluschke:2002}
S.~{Pl{\"u}schke}, M.~{Cervi{\~n}o}, R.~{Diehl}, K.~{Kretschmer}, D.~H.
  {Hartmann}, J.~{Kn{\"o}dlseder}, {Understanding $^{26}$Al Emission from
  Cygnus}, New Astronomy Review 46 (2002) 535--539.
\newblock \href {http://dx.doi.org/10.1016/S1387-6473(02)00197-5}
  {\path{doi:10.1016/S1387-6473(02)00197-5}}.

\bibitem{Diehl:2002}
R.~{Diehl}, {$^{26}$Al production in the Vela and Orion regions}, New Astronomy
  Review 46 (2002) 547--552.
\newblock \href {http://dx.doi.org/10.1016/S1387-6473(02)00199-9}
  {\path{doi:10.1016/S1387-6473(02)00199-9}}.

\bibitem{Dame:2001}
T.~M. {Dame}, D.~{Hartmann}, P.~{Thaddeus}, {The Milky Way in Molecular Clouds:
  A New Complete CO Survey}, \apj 547 (2001) 792--813.
\newblock \href {http://arxiv.org/abs/arXiv:astro-ph/0009217}
  {\path{arXiv:arXiv:astro-ph/0009217}}, \href
  {http://dx.doi.org/10.1086/318388} {\path{doi:10.1086/318388}}.

\bibitem{Knodlseder:1996}
J.~{Kn{\"o}dlseder}, {COMPTEL measurement of 1.8 MeV {\ensuremath{\gamma}}-ray
  line emission from radioactive $^{26}$Al}, in: Nuclear Astrophysics, 8th
  Workshop, 1996, p.~91.

\bibitem{Timmes:1997a}
F.~X. {Timmes}, R.~{Diehl}, D.~H. {Hartmann}, {Constraints from 26Al
  Measurements on the Galaxy's Recent Global Star Formation Rate and
  Core-Collapse Supernovae Rate}, \apj 479 (1997) 760--+.
\newblock \href {http://arxiv.org/abs/arXiv:astro-ph/9701242}
  {\path{arXiv:arXiv:astro-ph/9701242}}, \href
  {http://dx.doi.org/10.1086/303913} {\path{doi:10.1086/303913}}.

\bibitem{Diehl:2006d}
R.~{Diehl}, H.~{Halloin}, K.~{Kretschmer}, G.~G. {Lichti},
  V.~{Sch{\"o}nfelder}, A.~W. {Strong}, A.~{von Kienlin}, W.~{Wang}, P.~{Jean},
  J.~{Kn{\"o}dlseder}, J.-P. {Roques}, G.~{Weidenspointner}, S.~{Schanne},
  D.~H. {Hartmann}, C.~{Winkler}, C.~{Wunderer}, {Radioactive $^{26}$Al from
  massive stars in the Galaxy}, \nat 439 (2006) 45--47.
\newblock \href {http://arxiv.org/abs/arXiv:astro-ph/0601015}
  {\path{arXiv:arXiv:astro-ph/0601015}}, \href
  {http://dx.doi.org/10.1038/nature04364} {\path{doi:10.1038/nature04364}}.

\bibitem{Naya:1996}
J.~E. {Naya}, S.~D. {Barthelmy}, L.~M. {Bartlett}, N.~{Gehrels},
  M.~{Leventhal}, A.~{Parsons}, B.~J. {Teegarden}, J.~{Tueller}, {Detection of
  high-velocity $^{26}$Al towards the Galactic Centre}, \nat 384 (1996) 44--46.
\newblock \href {http://dx.doi.org/10.1038/384044a0}
  {\path{doi:10.1038/384044a0}}.

\bibitem{Chen:1997}
W.~{Chen}, R.~{Diehl}, N.~{Gehrels}, D.~{Hartmann}, M.~{Leising}, J.~E. {Naya},
  N.~{Prantzos}, J.~{Tueller}, P.~{von Ballmoos}, {Implications of the Broad
  $^{26}$Al 1809 keV Line Observed by GRIS}, in: C.~{Winkler}, T.~J.-L.
  {Courvoisier}, P.~{Durouchoux} (Eds.), The Transparent Universe, Vol. 382 of
  ESA Special Publication, 1997, pp. 105--+.

\bibitem{Sturner:1999}
S.~J. {Sturner}, J.~E. {Naya}, {On the Nature of the High-Velocity $^{26}$Al
  near the Galactic Center}, \apj 526 (1999) 200--206.
\newblock \href {http://dx.doi.org/10.1086/307979} {\path{doi:10.1086/307979}}.

\bibitem{Kretschmer:2013}
K.~{Kretschmer}, R.~{Diehl}, M.~{Krause}, A.~{Burkert}, K.~{Fierlinger},
  O.~{Gerhard}, J.~{Greiner}, W.~{Wang}, {Kinematics of massive star ejecta in
  the Milky Way as traced by $^{26}$Al}, \aap 559 (2013) A99.
\newblock \href {http://arxiv.org/abs/1309.4980} {\path{arXiv:1309.4980}},
  \href {http://dx.doi.org/10.1051/0004-6361/201322563}
  {\path{doi:10.1051/0004-6361/201322563}}.

\bibitem{Diehl:2006c}
R.~{Diehl}, H.~{Halloin}, K.~{Kretschmer}, A.~W. {Strong}, W.~{Wang},
  P.~{Jean}, G.~G. {Lichti}, J.~{Kn{\"o}dlseder}, J.-P. {Roques}, S.~{Schanne},
  V.~{Sch{\"o}nfelder}, A.~{von Kienlin}, G.~{Weidenspointner}, C.~{Winkler},
  C.~{Wunderer}, {$^{26}$Al in the inner Galaxy. Large-scale spectral
  characteristics derived with SPI/INTEGRAL}, \aap 449 (2006) 1025--1031.
\newblock \href {http://arxiv.org/abs/arXiv:astro-ph/0512334}
  {\path{arXiv:arXiv:astro-ph/0512334}}, \href
  {http://dx.doi.org/10.1051/0004-6361:20054301}
  {\path{doi:10.1051/0004-6361:20054301}}.

\bibitem{Krause:2015}
M.~G.~H. {Krause}, R.~{Diehl}, Y.~{Bagetakos}, E.~{Brinks}, A.~{Burkert},
  O.~{Gerhard}, J.~{Greiner}, K.~{Kretschmer}, T.~{Siegert}, {$^{26}$Al
  kinematics: superbubbles following the spiral arms?. Constraints from the
  statistics of star clusters and HI supershells}, \aap 578 (2015) A113.
\newblock \href {http://arxiv.org/abs/1504.03120} {\path{arXiv:1504.03120}},
  \href {http://dx.doi.org/10.1051/0004-6361/201525847}
  {\path{doi:10.1051/0004-6361/201525847}}.

\bibitem{Siegert:2017}
T.~{Siegert}, Positron-annihilation spectroscopy throughout the milky way,
  Ph.D. thesis, TU Munich (2017).

\bibitem{Martin:2009}
P.~{Martin}, J.~{Kn{\"o}dlseder}, R.~{Diehl}, G.~{Meynet}, {New estimates of
  the gamma-ray line emission of the Cygnus region from INTEGRAL/SPI
  observations}, \aap 506 (2009) 703--710.
\newblock \href {http://dx.doi.org/10.1051/0004-6361/200912178}
  {\path{doi:10.1051/0004-6361/200912178}}.

\bibitem{Diehl:2010}
R.~{Diehl}, M.~G. {Lang}, P.~{Martin}, H.~{Ohlendorf}, T.~{Preibisch},
  R.~{Voss}, P.~{Jean}, J.~{Roques}, P.~{von Ballmoos}, W.~{Wang}, {Radioactive
  $^{26}$Al from the Scorpius-Centaurus association}, \aap 522 (2010) A51+.
\newblock \href {http://arxiv.org/abs/1007.4462} {\path{arXiv:1007.4462}},
  \href {http://dx.doi.org/10.1051/0004-6361/201014302}
  {\path{doi:10.1051/0004-6361/201014302}}.

\bibitem{Diehl:2016b}
R.~{Diehl}, {New insights from cosmic gamma rays}, Journal of Physics
  Conference Series 703~(1) (2016) 012001.
\newblock \href {http://arxiv.org/abs/1602.09018} {\path{arXiv:1602.09018}},
  \href {http://dx.doi.org/10.1088/1742-6596/703/1/012001}
  {\path{doi:10.1088/1742-6596/703/1/012001}}.

 \bibitem{Pleintinger:2020}
M.~M.~M. {Pleintinger}, {Star groups and their nucleosynthesis}, Ph.D. Thesis, TU Munich (2020). 

\bibitem{Fujimoto:2018}
Y.~{Fujimoto}, M.~R. {Krumholz}, S.~{Tachibana}, {Short-lived radioisotopes in
  meteorites from Galactic-scale correlated star formation}, \mnras 480~(3)
  (2018) 4025--4039.
\newblock \href {http://arxiv.org/abs/1802.08695} {\path{arXiv:1802.08695}},
  \href {http://dx.doi.org/10.1093/mnras/sty2132}
  {\path{doi:10.1093/mnras/sty2132}}.
 
\bibitem{Pleintinger:2019}
M.~M.~M. {Pleintinger}, T.~{Siegert}, R.~{Diehl}, Y.~{Fujimoto}, J.~{Greiner},
  M.~G.~H. {Krause}, M.~R. {Krumholz}, {Comparing simulated $^{26}$Al maps to
  gamma-ray measurements}, \aap 632 (2019) A73.
\newblock \href {http://arxiv.org/abs/1910.06112} {\path{arXiv:1910.06112}},
  \href {http://dx.doi.org/10.1051/0004-6361/201935911}
  {\path{doi:10.1051/0004-6361/201935911}}.

\bibitem{Rodgers-Lee:2019}
D.~{Rodgers-Lee}, M.~G.~H. {Krause}, J.~{Dale}, R.~{Diehl}, {Synthetic
  $^{26}$Al emission from galactic-scale superbubble simulations}, \mnras
  490~(2) (2019) 1894--1912.
\newblock \href {http://arxiv.org/abs/1909.10978} {\path{arXiv:1909.10978}},
  \href {http://dx.doi.org/10.1093/mnras/stz2708}
  {\path{doi:10.1093/mnras/stz2708}}.

\bibitem{Chomiuk:2011}
L.~{Chomiuk}, M.~S. {Povich}, {Toward a Unification of Star Formation Rate
  Determinations in the Milky Way and Other Galaxies}, \aj 142 (2011) 197.
\newblock \href {http://arxiv.org/abs/1110.4105} {\path{arXiv:1110.4105}},
  \href {http://dx.doi.org/10.1088/0004-6256/142/6/197}
  {\path{doi:10.1088/0004-6256/142/6/197}}.

\bibitem{Diehl:2013b}
R.~{Diehl}, {Cosmic Gamma-Ray Spectroscopy}, Astronomical Review 8~(4) (2013)
  4--52.
\newblock \href {http://arxiv.org/abs/1307.4198} {\path{arXiv:1307.4198}}.

\bibitem{Janka:2007a}
H.~{Janka}, K.~{Langanke}, A.~{Marek}, G.~{Mart{\'{\i}}nez-Pinedo},
  B.~{M{\"u}ller}, {Theory of core-collapse supernovae}, \physrep 442 (2007)
  38--74.
\newblock \href {http://arxiv.org/abs/arXiv:astro-ph/0612072}
  {\path{arXiv:arXiv:astro-ph/0612072}}, \href
  {http://dx.doi.org/10.1016/j.physrep.2007.02.002}
  {\path{doi:10.1016/j.physrep.2007.02.002}}.

\bibitem{Crowther:2007}
P.~A. {Crowther}, {Physical Properties of Wolf-Rayet Stars}, \araa 45 (2007)
  177--219.
\newblock \href {http://arxiv.org/abs/arXiv:astro-ph/0610356}
  {\path{arXiv:arXiv:astro-ph/0610356}}, \href
  {http://dx.doi.org/10.1146/annurev.astro.45.051806.110615}
  {\path{doi:10.1146/annurev.astro.45.051806.110615}}.

\bibitem{Baumgartner:2013}
V.~{Baumgartner}, D.~{Breitschwerdt}, {Superbubble evolution in disk galaxies.
  I. Study of blow-out by analytical models}, \aap 557 (2013) A140.
\newblock \href {http://arxiv.org/abs/1402.0194} {\path{arXiv:1402.0194}},
  \href {http://dx.doi.org/10.1051/0004-6361/201321261}
  {\path{doi:10.1051/0004-6361/201321261}}.

\bibitem{Krause:2014}
M.~G.~H. {Krause}, R.~{Diehl}, {Dynamics and Energy Loss in Superbubbles},
  \apjl 794 (2014) L21.
\newblock \href {http://arxiv.org/abs/1409.7528} {\path{arXiv:1409.7528}},
  \href {http://dx.doi.org/10.1088/2041-8205/794/2/L21}
  {\path{doi:10.1088/2041-8205/794/2/L21}}.

\bibitem{Krause:2018}
M.~G.~H. {Krause}, A.~{Burkert}, R.~{Diehl}, K.~{Fierlinger}, B.~{Gaczkowski},
  D.~{Kroell}, J.~{Ngoumou}, V.~{Roccatagliata}, T.~{Siegert}, T.~{Preibisch},
  {Surround and Squash: the impact of superbubbles on the interstellar medium
  in Scorpius-Centaurus OB2}, \aap 619 (2018) A120.
\newblock \href {http://arxiv.org/abs/1808.04788} {\path{arXiv:1808.04788}},
  \href {http://dx.doi.org/10.1051/0004-6361/201732416}
  {\path{doi:10.1051/0004-6361/201732416}}.

\bibitem{Siegert:2017a}
T.~{Siegert}, R.~{Diehl}, {The $^{26}$Al Gamma-ray Line from Massive-Star
  Regions}, in: S.~{Kubono}, T.~{Kajino}, S.~{Nishimura}, T.~{Isobe},
  S.~{Nagataki}, T.~{Shima}, Y.~{Takeda} (Eds.), 14th International Symposium
  on Nuclei in the Cosmos (NIC2016), 2017, p. 020305.
\newblock \href {http://arxiv.org/abs/1609.08817} {\path{arXiv:1609.08817}},
  \href {http://dx.doi.org/10.7566/JPSCP.14.020305}
  {\path{doi:10.7566/JPSCP.14.020305}}.

\bibitem{Martin:2010b}
P.~{Martin}, J.~{Kn{\"o}dlseder}, G.~{Meynet}, R.~{Diehl}, {Predicted gamma-ray
  line emission from the Cygnus complex}, \aap 511 (2010) A86+.
\newblock \href {http://arxiv.org/abs/1001.1522} {\path{arXiv:1001.1522}},
  \href {http://dx.doi.org/10.1051/0004-6361/200912864}
  {\path{doi:10.1051/0004-6361/200912864}}.

\bibitem{Olano:1982}
C.~A. {Olano}, {On a model of local gas related to Gould's belt}, \aap 112
  (1982) 195--208.

\bibitem{Poppel:1997}
W.~{P\"oppel}, {The Gould Belt System and the Local Interstellar Medium},
  Fundamentals of Cosmic Physics 18 (1997) 1--271.

\bibitem{Poppel:2010}
W.~G.~L. {P{\"o}ppel}, E.~{Bajaja}, E.~M. {Arnal}, R.~{Morras}, {The
  interstellar medium surrounding the Scorpius-Centaurus association
  revisited}, \aap 512 (2010) A83+.
\newblock \href {http://dx.doi.org/10.1051/0004-6361/200811290}
  {\path{doi:10.1051/0004-6361/200811290}}.

\bibitem{Maiz-Apellaniz:2004}
J.~{Ma{\'{\i}}z-Apell{\'a}niz}, {Massive Young Clusters}, in: E.~J. {Alfaro},
  E.~{P{\'e}rez}, J.~{Franco} (Eds.), How Does the Galaxy Work?, Vol. 315 of
  Astrophysics and Space Science Library, 2004, p. 231.
\newblock \href {http://arxiv.org/abs/arXiv:astro-ph/0310162}
  {\path{arXiv:arXiv:astro-ph/0310162}}, \href
  {http://dx.doi.org/10.1007/1-4020-2620-X_46}
  {\path{doi:10.1007/1-4020-2620-X_46}}.

\bibitem{Perrot:2003}
C.~A. {Perrot}, I.~A. {Grenier}, {3D dynamical evolution of the interstellar
  gas in the Gould Belt}, \aap 404 (2003) 519--531.
\newblock \href {http://arxiv.org/abs/astro-ph/0303516}
  {\path{arXiv:astro-ph/0303516}}, \href
  {http://dx.doi.org/10.1051/0004-6361:20030477}
  {\path{doi:10.1051/0004-6361:20030477}}.

\bibitem{Knodlseder:2004a}
J.~{Kn{\"o}dlseder}, M.~{Valsesia}, M.~{Allain}, S.~{Boggs}, R.~{Diehl},
  P.~{Jean}, K.~{Kretschmer}, J.-P. {Roques}, V.~{Sch{\"o}nfelder},
  G.~{Vedrenne}, P.~{von Ballmoos}, G.~{Weidenspointner}, C.~{Winkler},
  {SPI/INTEGRAL Observation of 1809 keV Gamma-Ray Line Emission from the CYGNUS
  X Region}, in: V.~{Schoenfelder}, G.~{Lichti}, C.~{Winkler} (Eds.), 5th
  INTEGRAL Workshop on the INTEGRAL Universe, Vol. 552 of ESA Special
  Publication, 2004, pp. 33--+.

\bibitem{Maurin:2004}
D.~{Maurin}, S.~{Schanne}, P.~{Sizun}, D.~{Atti{\'e}}, S.~{Cordier},
  R.~{Diehl}, M.~{Gros}, P.~{Jean}, A.~{von Kienlin}, J.~{Kn{\"o}dlseder},
  {Search for $^{26}$Al Emission in the Vela Region with INTEGRAL/SPI}, in:
  V.~{Schoenfelder}, G.~{Lichti}, C.~{Winkler} (Eds.), 5th INTEGRAL Workshop on
  the INTEGRAL Universe, Vol. 552 of ESA Special Publication, 2004, pp. 107--+.

\bibitem{Voss:2012}
R.~{Voss}, P.~{Martin}, R.~{Diehl}, J.~S. {Vink}, D.~H. {Hartmann},
  T.~{Preibisch}, {Energetic feedback and $^{26}$Al from massive stars and
  their supernovae in the Carina region}, \aap 539 (2012) A66.
\newblock \href {http://arxiv.org/abs/1202.0282} {\path{arXiv:1202.0282}},
  \href {http://dx.doi.org/10.1051/0004-6361/201118209}
  {\path{doi:10.1051/0004-6361/201118209}}.

\bibitem{Reipurth:2008}
B.~{Reipurth}, N.~{Schneider}, {Star Formation and Young Clusters in Cygnus},
  Vol.~4, 2008, p.~36.

\bibitem{Diehl:1995}
R.~{Diehl}, {Imaging Diffuse Emission with COMPTEL}, Experimental Astronomy 6
  (1995) 103--108.
\newblock \href {http://dx.doi.org/10.1007/BF00419264}
  {\path{doi:10.1007/BF00419264}}.

\bibitem{Fierlinger:2012}
K.~M. {Fierlinger}, A.~{Burkert}, R.~{Diehl}, C.~{Dobbs}, D.~H. {Hartmann},
  M.~{Krause}, E.~{Ntormousi}, R.~{Voss}, {Molecular Cloud Disruption and
  Chemical Enrichment of the ISM Caused by Massive Star Feedback}, in:
  R.~{Capuzzo-Dolcetta}, M.~{Limongi}, A.~{Tornamb{\`e}} (Eds.), Advances in
  Computational Astrophysics: Methods, Tools, and Outcome, Vol. 453 of
  Astronomical Society of the Pacific Conference Series, 2012, p.~25.

\bibitem{Voss:2010a}
R.~{Voss}, R.~{Diehl}, J.~S. {Vink}, D.~H. {Hartmann}, {Probing the evolving
  massive star population in Orion with kinematic and radioactive tracers},
  \aap 520 (2010) A51+.
\newblock \href {http://arxiv.org/abs/1005.3827} {\path{arXiv:1005.3827}},
  \href {http://dx.doi.org/10.1051/0004-6361/201014408}
  {\path{doi:10.1051/0004-6361/201014408}}.

\bibitem{Comeron:2002}
F.~{Comer{\'o}n}, A.~{Pasquali}, G.~{Rodighiero}, V.~{Stanishev}, E.~{De
  Filippis}, B.~{L{\'o}pez Mart{\'{\i}}}, M.~C. {G{\'a}lvez Ortiz},
  A.~{Stankov}, R.~{Gredel}, {On the massive star contents of Cygnus OB2}, \aap
  389 (2002) 874--888.
\newblock \href {http://dx.doi.org/10.1051/0004-6361:20020648}
  {\path{doi:10.1051/0004-6361:20020648}}.

\bibitem{Knodlseder:2002}
J.~{Kn{\"o}dlseder}, M.~{Cervi{\~n}o}, J.-M. {Le Duigou}, G.~{Meynet},
  D.~{Schaerer}, P.~{von Ballmoos}, {Gamma-ray line emission from OB
  associations and young open clusters. II. The Cygnus region}, \aap 390 (2002)
  945--960.
\newblock \href {http://arxiv.org/abs/arXiv:astro-ph/0206045}
  {\path{arXiv:arXiv:astro-ph/0206045}}, \href
  {http://dx.doi.org/10.1051/0004-6361:20020799}
  {\path{doi:10.1051/0004-6361:20020799}}.

\bibitem{Cervino:2000}
M.~{Cervi{\~n}o}, J.~{Kn{\"o}dlseder}, D.~{Schaerer}, P.~{von Ballmoos},
  G.~{Meynet}, {Gamma-ray line emission from OB associations and young open
  clusters. I. Evolutionary synthesis models}, \aap 363 (2000) 970--983.
\newblock \href {http://arxiv.org/abs/arXiv:astro-ph/0010283}
  {\path{arXiv:arXiv:astro-ph/0010283}}.

\bibitem{Voss:2009}
R.~{Voss}, R.~{Diehl}, D.~H. {Hartmann}, M.~{Cervi{\~n}o}, J.~S. {Vink},
  G.~{Meynet}, M.~{Limongi}, A.~{Chieffi}, {Using population synthesis of
  massive stars to study the interstellar medium near OB associations}, \aap
  504 (2009) 531--542.
\newblock \href {http://arxiv.org/abs/0907.5209} {\path{arXiv:0907.5209}},
  \href {http://dx.doi.org/10.1051/0004-6361/200912260}
  {\path{doi:10.1051/0004-6361/200912260}}.

\bibitem{Kretschmer:2003a}
K.~{Kretschmer}, R.~{Diehl}, S.~{Pl{\"u}schke}, M.~{Cervi{\~n}o}, D.~H.
  {Hartmann}, {Radioactive 26Al in the Cygnus Region}, Astronomische
  Nachrichten Supplement 324 (2003) 76--+.

\bibitem{Brown:1994}
A.~G.~A. {Brown}, E.~J. {de Geus}, P.~T. {de Zeeuw}, {The Orion OB1
  association. 1: Stellar content}, \aap 289 (1994) 101--120.
\newblock \href {http://arxiv.org/abs/arXiv:astro-ph/9403051}
  {\path{arXiv:arXiv:astro-ph/9403051}}.

\bibitem{Brown:1995}
A.~G.~A. {Brown}, D.~{Hartmann}, W.~B. {Burton}, {The Orion OB1 association.
  II. The Orion-Eridanus Bubble.}, \aap 300 (1995) 903.
\newblock \href {http://arxiv.org/abs/astro-ph/9503016}
  {\path{arXiv:astro-ph/9503016}}.

\bibitem{Diehl:2003e}
R.~{Diehl}, K.~{Kretschmer}, S.~{Pl{\"u}schke}, M.~{Cervi{\~n}o}, D.~H.
  {Hartmann}, {Gamma-rays from massive stars in Cygnus and Orion}, in: K.~{van
  der Hucht}, A.~{Herrero}, C.~{Esteban} (Eds.), A Massive Star Odyssey: From
  Main Sequence to Supernova, Vol. 212 of IAU Symposium, 2003, pp. 706--+.

\bibitem{Fierlinger:2016}
K.~M. {Fierlinger}, A.~{Burkert}, E.~{Ntormousi}, P.~{Fierlinger},
  M.~{Schartmann}, A.~{Ballone}, M.~G.~H. {Krause}, R.~{Diehl}, {Stellar
  feedback efficiencies: supernovae versus stellar winds}, \mnras 456 (2016)
  710--730.
\newblock \href {http://arxiv.org/abs/1511.05151} {\path{arXiv:1511.05151}},
  \href {http://dx.doi.org/10.1093/mnras/stv2699}
  {\path{doi:10.1093/mnras/stv2699}}.

\bibitem{Preibisch:2008}
T.~{Preibisch}, E.~{Mamajek}, {The Nearest OB Association: Scorpius-Centaurus
  (Sco OB2)}, 2008, p. 235.

\bibitem{de-Geus:1989}
E.~J. {de Geus}, P.~T. {de Zeeuw}, J.~{Lub}, {Physical parameters of stars in
  the Scorpio-Centaurus OB association}, \aap 216 (1989) 44--61.

\bibitem{de-Geus:1992}
E.~J. {de Geus}, {Interactions of stars and interstellar matter in Scorpio
  Centaurus}, \aap 262 (1992) 258--270.

\bibitem{Rizzuto:2016}
A.~C. {Rizzuto}, M.~J. {Ireland}, T.~J. {Dupuy}, A.~L. {Kraus}, {Dynamical
  Masses of Young Stars. I. Discordant Model Ages of Upper Scorpius}, \apj 817
  (2016) 164.
\newblock \href {http://arxiv.org/abs/1512.05371} {\path{arXiv:1512.05371}},
  \href {http://dx.doi.org/10.3847/0004-637X/817/2/164}
  {\path{doi:10.3847/0004-637X/817/2/164}}.

\bibitem{Winkel:2008}
B.~{Winkel}, J.~{Kerp}, P.~M.~W. {Kalberla}, R.~{Keller}, {The Effelsberg-Bonn
  HI Survey (EBHIS)}, in: {R.~Minchin \& E.~Momjian} (Ed.), The Evolution of
  Galaxies Through the Neutral Hydrogen Window, Vol. 1035 of American Institute
  of Physics Conference Series, 2008, pp. 259--261.
\newblock \href {http://arxiv.org/abs/0708.1858} {\path{arXiv:0708.1858}},
  \href {http://dx.doi.org/10.1063/1.2973598} {\path{doi:10.1063/1.2973598}}.

\bibitem{Gaczkowski:2017}
B.~{Gaczkowski}, V.~{Roccatagliata}, S.~{Flaischlen}, D.~{Kr{\"o}ll}, M.~G.~H.
  {Krause}, A.~{Burkert}, R.~{Diehl}, K.~{Fierlinger}, J.~{Ngoumou},
  T.~{Preibisch}, {Squeezed between shells? The origin of the Lupus I molecular
  cloud. II. APEX CO and GASS H I observations}, \aap 608 (2017) A102.
\newblock \href {http://arxiv.org/abs/1710.07446} {\path{arXiv:1710.07446}},
  \href {http://dx.doi.org/10.1051/0004-6361/201628508}
  {\path{doi:10.1051/0004-6361/201628508}}.

\bibitem{Preibisch:1999}
T.~{Preibisch}, H.~{Zinnecker}, {The History of Low-Mass Star Formation in the
  Upper Scorpius OB Association}, \aj 117 (1999) 2381--2397.
\newblock \href {http://dx.doi.org/10.1086/300842} {\path{doi:10.1086/300842}}.

\bibitem{Zinnecker:2007}
H.~{Zinnecker}, H.~W. {Yorke}, {Toward Understanding Massive Star Formation},
  \araa 45 (2007) 481--563.
\newblock \href {http://arxiv.org/abs/0707.1279} {\path{arXiv:0707.1279}},
  \href {http://dx.doi.org/10.1146/annurev.astro.44.051905.092549}
  {\path{doi:10.1146/annurev.astro.44.051905.092549}}.

\bibitem{Lee:2009}
H.~{Lee}, W.~P. {Chen}, {Triggered Star Formation on the Border of the
  Orion-Eridanus Superbubble}, \apj 694 (2009) 1423--1434.
\newblock \href {http://arxiv.org/abs/arXiv:astro-ph/0608216}
  {\path{arXiv:arXiv:astro-ph/0608216}}, \href
  {http://dx.doi.org/10.1088/0004-637X/694/2/1423}
  {\path{doi:10.1088/0004-637X/694/2/1423}}.

\bibitem{Churazov:2020}
E.~{Churazov}, L.~{Bouchet}, P.~{Jean}, E.~{Jourdain}, J.~{Kn{\"o}dlseder},
  R.~{Krivonos}, J.-P. {Roques}, S.~{Sazonov}, T.~{Siegert}, A.~{Strong},
  R.~{Sunyaev}, {INTEGRAL results on the electron-positron annihilation
  radiation and X-ray \& Gamma-ray diffuse emission of the Milky Way}, \nar 90
  (2020) 101548.
\newblock \href {http://dx.doi.org/10.1016/j.newar.2020.101548}
  {\path{doi:10.1016/j.newar.2020.101548}}.

\bibitem{Prantzos:2011}
N.~{Prantzos}, C.~{Boehm}, A.~M. {Bykov}, R.~{Diehl}, K.~{Ferri{\`e}re},
  N.~{Guessoum}, P.~{Jean}, J.~{Knoedlseder}, A.~{Marcowith}, I.~V.
  {Moskalenko}, A.~{Strong}, G.~{Weidenspointner}, {The 511 keV emission from
  positron annihilation in the Galaxy}, Reviews of Modern Physics 83 (2011)
  1001--1056.
\newblock \href {http://arxiv.org/abs/1009.4620} {\path{arXiv:1009.4620}},
  \href {http://dx.doi.org/10.1103/RevModPhys.83.1001}
  {\path{doi:10.1103/RevModPhys.83.1001}}.

\bibitem{Siegert:2016}
T.~{Siegert}, R.~{Diehl}, G.~{Khachatryan}, M.~G.~H. {Krause},
  F.~{Guglielmetti}, J.~{Greiner}, A.~W. {Strong}, X.~{Zhang}, {Gamma-ray
  spectroscopy of positron annihilation in the Milky Way}, \aap 586 (2016) A84.
\newblock \href {http://arxiv.org/abs/1512.00325} {\path{arXiv:1512.00325}},
  \href {http://dx.doi.org/10.1051/0004-6361/201527510}
  {\path{doi:10.1051/0004-6361/201527510}}.

\bibitem{Knodlseder:2003}
J.~{Kn{\"o}dlseder}, V.~{Lonjou}, P.~{Jean}, M.~{Allain}, P.~{Mandrou}, J.-P.
  {Roques}, G.~K. {Skinner}, G.~{Vedrenne}, P.~{von Ballmoos},
  G.~{Weidenspointner}, P.~{Caraveo}, B.~{Cordier}, V.~{Sch{\"o}nfelder}, B.~J.
  {Teegarden}, {Early SPI/INTEGRAL constraints on the morphology of the 511 keV
  line emission in the 4th galactic quadrant}, \aap 411 (2003) L457--L460.
\newblock \href {http://arxiv.org/abs/arXiv:astro-ph/0309442}
  {\path{arXiv:arXiv:astro-ph/0309442}}, \href
  {http://dx.doi.org/10.1051/0004-6361:20031437}
  {\path{doi:10.1051/0004-6361:20031437}}.

\bibitem{Rugel:2009}
G.~{Rugel}, T.~{Faestermann}, K.~{Knie}, G.~{Korschinek}, M.~{Poutivtsev},
  D.~{Schumann}, N.~{Kivel}, I.~{G{\"u}nther-Leopold}, R.~{Weinreich},
  M.~{Wohlmuther}, {New Measurement of the Fe60 Half-Life}, Physical Review
  Letters 103~(7) (2009) 072502--+.
\newblock \href {http://dx.doi.org/10.1103/PhysRevLett.103.072502}
  {\path{doi:10.1103/PhysRevLett.103.072502}}.

\bibitem{Wallner:2015}
A.~{Wallner}, T.~{Faestermann}, J.~{Feige}, C.~{Feldstein}, K.~{Knie},
  G.~{Korschinek}, W.~{Kutschera}, A.~{Ofan}, M.~{Paul}, F.~{Quinto},
  G.~{Rugel}, P.~{Steier}, {Abundance of live $^{244}$Pu in deep-sea reservoirs
  on Earth points to rarity of actinide nucleosynthesis}, Nature Communications
  6 (2015) 5956.
\newblock \href {http://arxiv.org/abs/1509.08054} {\path{arXiv:1509.08054}},
  \href {http://dx.doi.org/10.1038/ncomms6956} {\path{doi:10.1038/ncomms6956}}.

\bibitem{Limongi:2006}
M.~{Limongi}, A.~{Chieffi}, {Nucleosynthesis of $^{60}$Fe in massive stars},
  New Astronomy Review 50 (2006) 474--476.
\newblock \href {http://arxiv.org/abs/arXiv:astro-ph/0512598}
  {\path{arXiv:arXiv:astro-ph/0512598}}, \href
  {http://dx.doi.org/10.1016/j.newar.2006.06.005}
  {\path{doi:10.1016/j.newar.2006.06.005}}.

\bibitem{Limongi:2006a}
M.~{Limongi}, A.~{Chieffi}, {The Nucleosynthesis of $^{26}$Al and $^{60}$Fe in
  Solar Metallicity Stars Extending in Mass from 11 to 120 M$_{\odot}$: The
  Hydrostatic and Explosive Contributions}, \apj 647 (2006) 483--500.
\newblock \href {http://arxiv.org/abs/arXiv:astro-ph/0604297}
  {\path{arXiv:arXiv:astro-ph/0604297}}, \href
  {http://dx.doi.org/10.1086/505164} {\path{doi:10.1086/505164}}.

\bibitem{Woosley:1997}
S.~E. {Woosley}, {Neutron-rich Nucleosynthesis in Carbon Deflagration
  Supernovae}, \apj 476 (1997) 801--+.
\newblock \href {http://dx.doi.org/10.1086/303650} {\path{doi:10.1086/303650}}.

\bibitem{Woosley:2007b}
S.~E. {Woosley}, A.~{Heger}, {Nucleosynthesis and remnants in massive stars of
  solar metallicity}, \physrep 442 (2007) 269--283.
\newblock \href {http://arxiv.org/abs/arXiv:astro-ph/0702176}
  {\path{arXiv:arXiv:astro-ph/0702176}}, \href
  {http://dx.doi.org/10.1016/j.physrep.2007.02.009}
  {\path{doi:10.1016/j.physrep.2007.02.009}}.

\bibitem{Smith:2004}
D.~M. {Smith}, {Gamma-Ray Line Observations with RHESSI}, in:
  V.~{Schoenfelder}, G.~{Lichti}, C.~{Winkler} (Eds.), 5th INTEGRAL Workshop on
  the INTEGRAL Universe, Vol. 552 of ESA Special Publication, 2004, pp. 45--+.

\bibitem{Wang:2007}
W.~Wang, Study of long-lived radioactive sources in the galaxy with
  integral/spi, Ph d thesis, TU Munich, Munich, Germany (September 2007).

\bibitem{Diehl:2018}
R.~{Diehl}, T.~{Siegert}, J.~{Greiner}, M.~{Krause}, K.~{Kretschmer},
  M.~{Lang}, M.~{Pleintinger}, A.~W. {Strong}, C.~{Weinberger}, X.~{Zhang},
  {INTEGRAL/SPI {$\gamma$}-ray line spectroscopy. Response and background
  characteristics}, \aap 611 (2018) A12.
\newblock \href {http://dx.doi.org/10.1051/0004-6361/201731815}
  {\path{doi:10.1051/0004-6361/201731815}}.

\bibitem{Wang:2020}
W.~{Wang}, T.~{Siegert}, Z.~G. {Dai}, R.~{Diehl}, J.~{Greiner}, A.~{Heger},
  M.~{Krause}, M.~{Lang}, M.~M.~M. {Pleintinger}, X.~L. {Zhang}, {Gamma-Ray
  Emission of $^{60}$Fe and $^{26}$Al Radioactivity in Our Galaxy}, \apj
  889~(2) (2020) 169.
\newblock \href {http://arxiv.org/abs/1912.07874} {\path{arXiv:1912.07874}},
  \href {http://dx.doi.org/10.3847/1538-4357/ab6336}
  {\path{doi:10.3847/1538-4357/ab6336}}.

\bibitem{Kachelries:2015}
M.~{Kachelrie{\ss}}, A.~{Neronov}, D.~V. {Semikoz}, {Signatures of a Two
  Million Year Old Supernova in the Spectra of Cosmic Ray Protons, Antiprotons,
  and Positrons}, Physical Review Letters 115~(18) (2015) 181103.
\newblock \href {http://arxiv.org/abs/1504.06472} {\path{arXiv:1504.06472}},
  \href {http://dx.doi.org/10.1103/PhysRevLett.115.181103}
  {\path{doi:10.1103/PhysRevLett.115.181103}}.

\bibitem{Wallner:2016}
A.~{Wallner}, J.~{Feige}, N.~{Kinoshita}, M.~{Paul}, L.~K. {Fifield},
  R.~{Golser}, M.~{Honda}, U.~{Linnemann}, H.~{Matsuzaki}, S.~{Merchel},
  G.~{Rugel}, S.~G. {Tims}, P.~{Steier}, T.~{Yamagata}, S.~R. {Winkler},
  {Recent near-Earth supernovae probed by global deposition of interstellar
  radioactive $^{60}$Fe}, \nat 532 (2016) 69--72.
\newblock \href {http://dx.doi.org/10.1038/nature17196}
  {\path{doi:10.1038/nature17196}}.

\end{thebibliography}

\end{document}